# Coherence-Based Identification of Carbon-Based Spin Qubits in Hexagonal Boron Nitride from First Principles

Hyeonsu Kim[1], Jaewook Lee[1], Huijin Park[1], and Hosung Seo[1,2*]

[1]SKKU Advanced Institute of Nanotechnology, Sungkyunkwan University, Suwon, Gyeonggi 16419, Republic of Korea

[2]Department of Quantum Information Engineering, Sungkyunkwan University, Suwon, 16419, Republic of Korea

(*correspondence to seo.hosung@skku.edu)

## Abstract

Carbon-related defects in hexagonal boron nitride are promising room-temperature single-spin qubits and quantum sensors, but their atomic structures remain largely unidentified. Here we show, using first-principles calculations of electron-spin decoherence, that the atomic structure of each defect is imprinted in its spin coherence. Mapping the Hahn-echo dynamics of seven candidate carbon defects across magnetic field and four isotope-engineered nuclear-spin baths, we find that electron-spin-echo envelope modulation emerges at defect-specific magnetic fields, at which the nearest-neighbor nuclear spins satisfy a cancellation condition set by their hyperfine and quadrupole couplings. Both the fields and the modulation frequencies follow from an analytical model using computed hyperfine and quadrupole tensors alone, and they shift or vanish upon isotope substitution. At low fields, the field dependence of the coherence time separates the defects into two classes according to the sublattice occupied by carbon. These decoherence fingerprints, directly testable in isotope-engineered samples, establish a structural identification route complementary to optical spectroscopy.

## Introduction

Solid-state spin defects in semiconductors have become powerful platforms for quantum sensing [1–5], delivering nanoscale magnetometry [6,7], coherent control of electron-nuclear spin registers [8,9], and millisecond coherence in isotopically purified samples [10,11]. The nitrogen-vacancy (NV) center in diamond now enables applications spanning nanoscale nuclear magnetic resonance with chemical resolution [12], magnetic imaging of condensed-matter systems [13], and biological sensing in cells [14]. Yet realizing the full sensing potential of these platforms requires placing the qubit within sub-nanometer proximity of the target, a condition that is difficult to meet in bulk crystal hosts even after extensive near-surface engineering [15–17]. These challenges motivate the search for spin defects in two-dimensional (2D) materials [18], where atomic-scale thickness naturally places the qubit within angstroms of any adjacent surface or interface.

Hexagonal boron nitride (h-BN) is a leading example of this 2D route, hosting optically addressable spin defects within atomically thin layers that integrate into van der Waals heterostructures [19–21] and photonic devices such as waveguides and cavities [22–24]. The negatively charged boron vacancy ($V_B^-$) was instrumental in establishing h-BN as a viable spin-qubit platform through room-temperature optically detected magnetic resonance (ODMR) and coherent control [25–30]. $V_B^-$ ensembles now serve as wide-field magnetometers for 2D materials [31,32] and as multimodal sensors of temperature, pressure, strain, and electric field [33,34].

However, its weak photoluminescence and ensemble-level ODMR contrast (typically a few percent) have hindered both single-defect optical readout and single-spin sensitivity [25,30]. Carbon-related defects have emerged as promising candidates to fill this gap, and are now widely regarded as the origin of many bright spin-active optical signatures in h-BN [21,35–38]. Subsequent experiments have demonstrated room-temperature single-spin ODMR in carbon-related defects [39,40], alongside coherent control of individual nuclear spins [41] and vectorial and multi-species magnetometry that establishes these defects as versatile nanoscale sensors [42–44]. Together, these results establish carbon-related defects as a single-spin sensing platform complementary to $V_B^-$ ensembles.

Despite this rapid progress, the atomic structure of most carbon-related spin defects in h-BN remains poorly understood [35,45–47]. Carbon can enter the h-BN lattice as substitutional monomers [36], nearest-neighbor donor-acceptor pairs [41], trimers [48,49], and tetramers [50], with further candidates including split interstitials, larger carbon clusters, and carbon-vacancy complexes [51]. Yet narrowing this candidate space by optical means has proved difficult. Photoluminescence signatures of different carbon defects often overlap, and their ODMR spectra rarely resolve defect-specific hyperfine structure [36,52]. Thus, conclusive identification of carbon-related defects demands joint theory-experiment efforts to establish spectroscopic fingerprints that are both specific to atomic structure and complementary to optical signatures [35].

Spin decoherence encodes the nuclear-spin-bath structure and hyperfine environment of the defect itself. In h-BN, every native boron and nitrogen isotope possesses a non-zero nuclear spin, so any embedded qubit is coupled to a dense nuclear spin bath whose dynamics depends on isotope composition, external magnetic field, and the local hyperfine structure of the defect [53,54]. For $V_B^-$, this dependence has recently been elucidated [54]: electron spin echo envelope modulation (ESEEM) arises from individual bath nuclei, with frequencies set by their hyperfine and quadrupole couplings. By extension, each carbon-related defect's distinct hyperfine environment should imprint its distinct hyperfine environment on the spin dynamics, offering a route to identification through coherence properties. However, such a systematic, decoherence-based comparison across carbon-defect candidates remains absent for h-BN [35].

Here we present a systematic first-principles investigation of the electron-spin decoherence of seven representative carbon defects in h-BN, spanning monomers to tetramers. Combining hybrid-functional density functional theory with the cluster correlation expansion [55–58], we map their Hahn-echo and Carr-Purcell-Meiboom-Gill (CPMG) coherence dynamics across magnetic field and four isotope-engineered nuclear spin baths ($^{10}B^{15}N$, $^{11}B^{15}N$, $^{10}B^{14}N$, $^{11}B^{14}N$). We find that each defect imprints its atomic structure on its own coherence: above a bath-specific magnetic-field transition boundary (TB), ESEEM emerges at sharply defined magnetic fields set by strongly coupled nuclear spins adjacent to the defect. Below the transition boundary, the magnetic-field dependence of $T_2$ sorts the defects into two classes according to the sublattice their carbon atoms occupy. Together, these signatures establish spin decoherence as a structural fingerprint of carbon defects in h-BN, leading to a defect-identification framework complementary to optical spectroscopy and directly testable in currently available isotope-engineered h-BN samples.

## Results

**Electronic structure and spin Hamiltonian.** As summarized in Fig. 1, we consider seven carbon-related defects, classified into two groups according to the sublattice site that carries the spin density: N-site defects ($C_N$, $C_2C_N$, $C_B(C_N)_3$, $C_NC_B^+$-DAP2) and B-site defects ($C_B$, $C_2C_B$, $C_N(C_B)_3$). The substitutional monomers $C_N$ and $C_B$ have been studied as paramagnetic carbon centers [36,46] and serve as building blocks of the more complex defects considered here. The dimer defect $C_NC_B^+$-DAP2 has recently attracted attention as a single-photon emitter (SPE) and spin-qubit candidate [36,41], and the trimers $C_2C_N$ and $C_2C_B$ are widely discussed SPE candidates [48,49,52]. The tetramers $C_B(C_N)_3$ and $C_N(C_B)_3$ have been theoretically proposed as optically addressable spin-triplet qubits, but their experimental realization has not yet been reported [50,59].

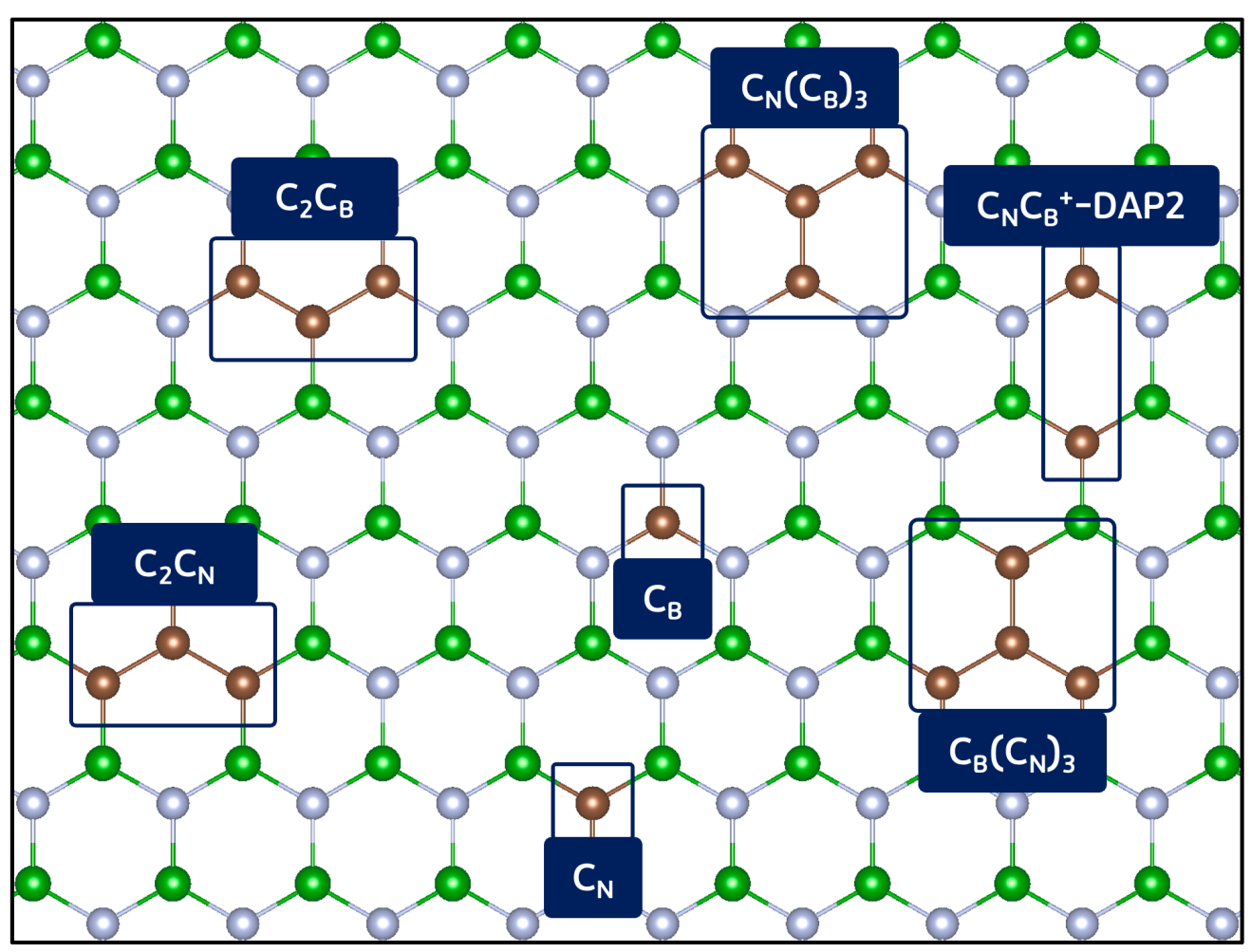

**Figure 1. Crystal structures of the seven carbon-related defects in h-BN.** Brown spheres are carbon, green spheres boron, and light gray spheres nitrogen. The defects are the substitutional monomers $C_N$ and $C_B$, the dimer defect $C_NC_B^+$-DAP2, the trimer defects $C_2C_N$ (= $C_NC_BC_N$) and $C_2C_B$ (= $C_BC_NC_B$), and the tetramer defects $C_N(C_B)_3$ and $C_B(C_N)_3$.

The electronic structure and spin Hamiltonian parameters of all seven defects were computed using hybrid-functional DFT (see Methods), and the resulting hyperfine and quadrupole tensors of the atoms surrounding each defect are tabulated in Supplementary Tables S1–S14. In all cases the spin density is predominantly localized on the carbon atoms, so each defect spin couples most strongly to its nearest-neighbor (NN) nuclear spins through the Fermi contact interaction. Because the nearest neighbors of a nitrogen site are boron atoms, N-site defects couple mainly to NN boron nuclei, and B-site defects to NN nitrogen nuclei. This distinction underlies the coherence differences developed below. In $C_NC_B^+$-DAP2 the spin density localizes predominantly near the $C_N$ site [41], and this defect is therefore assigned to the N-site group.

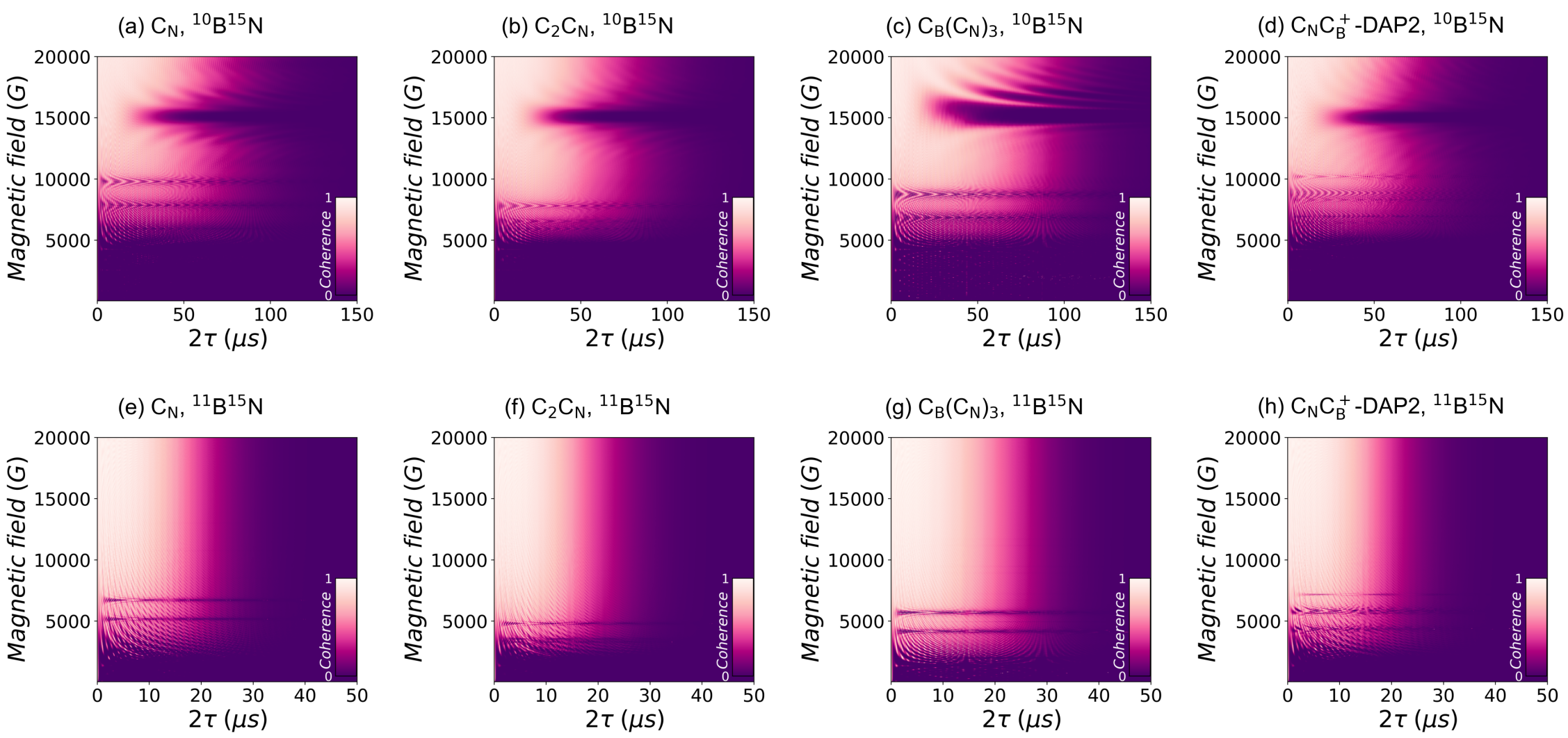

**Figure 2.** Hahn-echo coherence colormaps as a function of free evolution time (2τ) and external magnetic field (50–20000 G) for four representative N-site defects in two isotope baths. Row 1 (a–d) is the $^{10}B^{15}N$ bath (time axis up to 150 μs). Row 2 (e–h) is the $^{11}B^{15}N$ bath (time axis up to 50 μs). (a, e) $C_N$, (b, f) $C_2C_N$, (c, g) $C_B(C_N)_3$, (d, h) $C_NC_B^+$-DAP2.

**Coherence map and the transition boundary.** Figure 2 presents the Hahn-echo coherence of the four N-site defects ($C_N$, $C_2C_N$, $C_B(C_N)_3$, and $C_NC_B^+$-DAP2) as maps over the free-evolution time 2τ and the external magnetic field from 50 to 20,000 G, computed in the $^{10}B^{15}N$ (Fig. 2a–d) and $^{11}B^{15}N$ (Fig. 2e–h) baths; corresponding maps of the three B-site defects are shown in Supplementary Fig. S1. Three features stand out in Fig. 2. First, in both baths the coherence undergoes pronounced dephasing at low field and then recovers, with the onset of recovery near 5,000 G in the $^{10}B^{15}N$ bath and near 2,000 G in the $^{11}B^{15}N$ bath. Second, strong ESEEM patterns emerge at specific magnetic fields whose positions and shapes are defect dependent; $C_N$, for example, shows modulation near 8,000 and 10,000 G (Fig. 2a). Third, in the $^{10}B^{15}N$ bath a broad decoherence band appears near 15,000 G for all four defects, but no such band appears in the $^{11}B^{15}N$ bath (Fig. 2e–h). We analyze the first feature in this subsection and the second and third in the subsections that follow.

The boundary between the two regimes can be understood from our previous study of $V_B^-$ spin decoherence in h-BN [54]. For a spin qubit in the dense nuclear spin bath of h-BN, the dominant decoherence mechanism changes at a characteristic field strength, which we refer to as the transition boundary (TB); it lies near 5,000 G in $^{10}B$-rich baths and near 2,000 G in $^{11}B$-rich baths [54], consistent with Figure 2. Below the TB, each bath nuclear spin individually modulates the central-spin coherence, and the modulation is strongly enhanced whenever a nuclear spin satisfies the cancellation condition, a resonance-like criterion at which the nuclear Zeeman splitting compensates the local hyperfine and quadrupole splittings. Since different nuclear spins meet this condition at different fields, a substantial fraction of the bath spins with $I \geq 1$ collectively drive strong dephasing across the entire low-field regime. Above the TB these single-nucleus modulation effects are suppressed, and pairwise flip-flop dynamics among bath nuclear spins become the primary decoherence source. The TB is thus set by the bath spin properties rather than by the defect identity.

**Coherence modulations above TB.** We now turn to the second feature of Fig. 2, the chevron-shaped modulation patterns that appear above the TB. The fields at which they appear differ from defect to defect and shift when the isotope surrounding the defect is changed (e.g., $^{10}B$ to $^{11}B$); the pattern is therefore a structural fingerprint carried by the coherence itself. We first trace the mechanism for $C_N$ in the $^{10}B^{15}N$ bath (Fig. 2a) and then extend the analysis to the remaining defects.

Figure 3a shows the Hahn-echo coherence of $C_N$ in the $^{10}B^{15}N$ bath at $B_0$ = 9800 G. The coherence decays as $\exp(-(t/T_2)^n)$ with $T_2$ = 71.3 μs and n = 2.23, while exhibiting strong ESEEM at two main frequencies, 0.13 and 5.46 MHz (Fourier spectrum in the inset). To identify the origin of the modulation, Fig. 3b shows the coherence computed with the bath reduced to the three NN $^{10}B$ nuclear spins, which are related by the local symmetry and modulate the coherence identically, multiplied by the same decay envelope (the reduced bath alone produces no decay). Its Fourier spectrum (inset) reproduces both the 0.13 MHz peak and the high-frequency triplet near 5.46 MHz of the full-bath result. The periodic beating thus arises from the NN $^{10}B$ nuclear spins, while the remaining bath spins contribute the overall decay through spin-pair flip-flop dynamics.

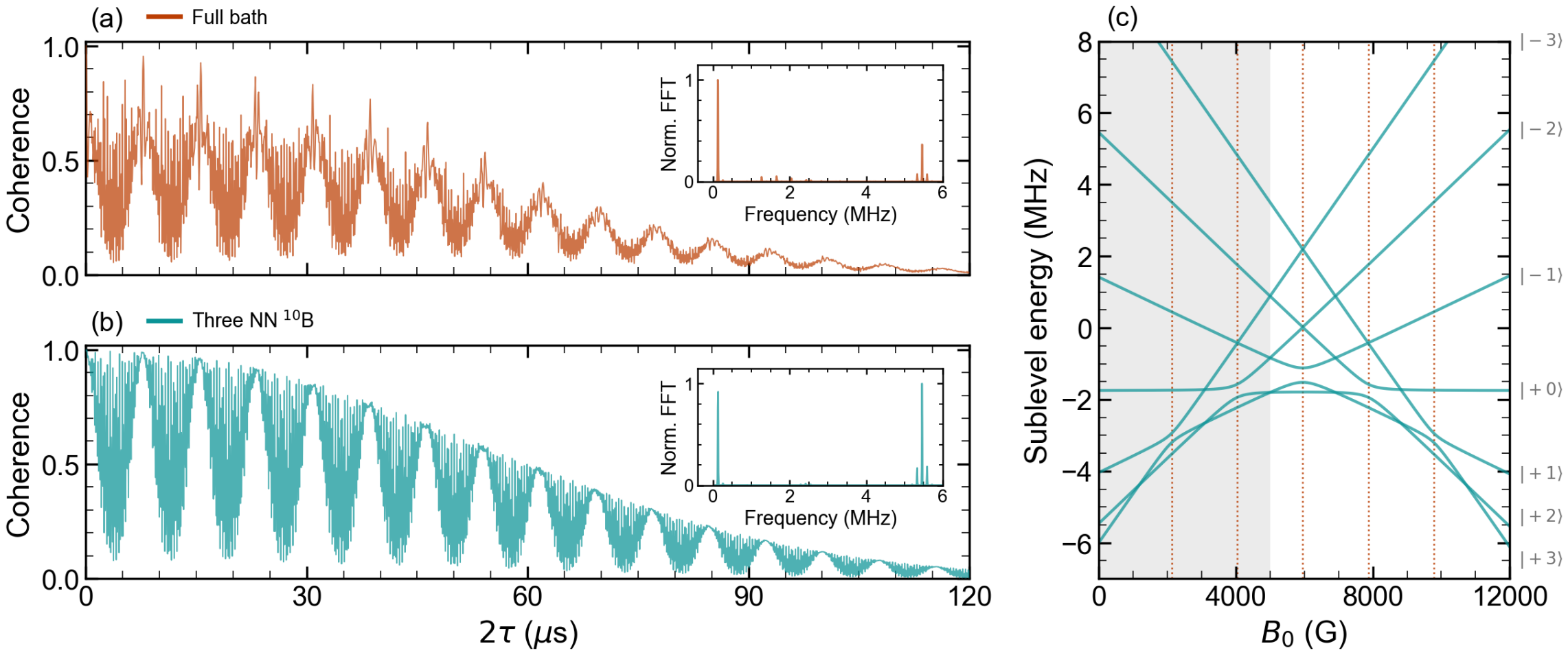

**Figure 3. Microscopic origin of the modulation pattern of the $C_N$ defect in the $^{10}B^{15}N$ bath.** Hahn-echo coherence function at $B_0$ = 9800 G for the full nuclear-spin bath (a) and for the NN $^{10}B$ nuclear spins (b), the latter multiplied by the decay envelope, $\exp(-(t/T_2)^n)$ with $T_2$ = 71.3 μs and n = 2.23. Insets: Fourier spectra of each coherence function. (c) Nuclear sublevel energies of the NN $^{10}B$ (I = 3) in the $m_S = -1/2$ qubit manifold. Orange dotted lines mark the five fields at which two sublevels separated by $\Delta m_I = 2$ satisfy the cancellation condition: 2150, 4050, 5950, 7850, and 9800 G. The shaded region lies below the transition boundary of the $^{10}B^{15}N$ bath, where individual ESEEM features are not resolved. Labels on the right give the $m_I$ from which each branch is predominantly built.

To trace the origin of these modulations, Figure 3c shows the nuclear sublevel energies of the NN $^{10}B$ (I = 3) in the $m_S = -1/2$ qubit manifold as a function of $B_0$. At zero field the seven sublevels are split on the megahertz scale by the hyperfine interaction ($A_{zz}$ = −5.45 MHz), which acts as an effective field of 2.7 MHz along the qubit quantization axis and separates the sublevels in proportion to $m_I$, and by the quadrupole interaction ($C_Q$ = 8.76 MHz), which shifts them in proportion to $m_I^2$. As $B_0$ increases, the nuclear Zeeman term progressively cancels this effective hyperfine field, so sublevels of different $m_I$ approach one another; because the quadrupole asymmetry mixes states differing by $\Delta m_I = 2$, the crossings open into anti-crossings, at which the cancellation condition is met and the ESEEM modulation depth becomes maximal [54]. Five such anti-crossings occur, evenly spaced in field. The two lowest lie below the TB, where the coherence decays within a few hundred nanoseconds and no modulation survives; the third, just above the boundary, is not sharply resolved; the two highest produce the pronounced modulations of Fig. 2a, including the one at 9800 G analyzed here.

This also explains why nuclear modulation, generally suppressed above the TB, reappears here, and why only the nearest neighbors carry it. The electric field gradient is essentially a bulk property, so the quadrupole interaction is nearly identical at every boron site, and it is the hyperfine coupling that sets the cancellation fields: the nuclear Zeeman splitting reaches the megahertz scale only at several thousand gauss, so the larger the hyperfine coupling, the higher the cancellation fields lie. Only the NN $^{10}B$ nuclei, with by far the largest Fermi-contact coupling, meet the condition above the 5000 G TB of the $^{10}B$ baths. All other bath nuclei either meet it below the TB, where the coherence has already decayed, or never meet it at all.

Near each cancellation field only the two anti-crossing sublevels carry the dynamics, so the qubit and its NN nucleus reduce to a two-level pseudo-spin (an effective S = 1/2 coupled to an effective I = 1/2), for which the Hahn-echo coherence takes the analytic form $1 - 2k\sin^2(\omega_-\tau/2)\sin^2(\omega_+\tau/2))$, with modulation depth $k$ and frequencies $\omega_\pm$ set by the splittings of the sublevel pair in the two qubit spin states (see Supplementary Note §S3 for detailed derivation). In the qubit state that meets the cancellation condition, the Zeeman, hyperfine, and quadrupole contributions cancel one another, leaving the anti-crossing gap of Fig. 3c as $\omega_-$; in the other state they add, and the same pair remains widely separated, giving $\omega_+$. Evaluating the model with the calculated hyperfine and quadrupole tensors alone yields $\omega_- = 0.26$ MHz and $\omega_+ = 10.92$ MHz at 9800 G. Since the coherence is followed as a function of the total evolution time $2\tau$, the modulation appears at $\omega_-/2 = 0.13$ MHz and $\omega_+/2 = 5.46$ MHz, the latter flanked by sum and difference sidebands at $(\omega_+ \pm \omega_-)/2 = 5.33$ and 5.59 MHz, precisely the features observed in Fig. 3b.

**Defect and isotope dependence of the modulation pattern.** Having established the mechanism for $C_N$, we now examine how the modulation pattern varies across defects. Figures 2b–d show the coherence maps of the other three N-site defects in the $^{10}B^{15}N$ bath. Each carries several modulation bands above the TB, at defect-specific fields: near 5550, 6550, and 7800 G for $C_2C_N$; near 6850 and 8750 G for $C_B(C_N)_3$; and near 6950, 8300, 8900, and 10,200 G for $C_NC_B^+$-DAP2. The cancellation condition accounts for these positions as well: a defect can couple strongly to several neighboring nuclei with different hyperfine couplings, and each meets the condition at its own set of fields (see Supplementary Figs. S2–S7 for the FFT spectra).

In the $^{11}B^{15}N$ bath (Fig. 2e–h), modulation bands again appear above the TB, at 5150 and 6700 G for $C_N$, 3300 and 4800 G for $C_2C_N$, 4200 and 5700 G for $C_B(C_N)_3$, and 4400, 5650, and 7150 G for $C_NC_B^+$-DAP2. Every field has shifted with respect to the $^{10}B^{15}N$ maps and the number of bands is reduced, as expected from the isotope change: the $^{11}B$ gyromagnetic ratio is three times that of $^{10}B$, and its nuclear spin I = 3/2 allows only two $\Delta m_I = 2$ transitions instead of five. The same pseudo-spin model reproduces these changes. In addition, the TB itself drops to about 2000 G, so bands can be resolved at fields below the boundary of the $^{10}B$ baths, as for the 4200 G band of $C_B(C_N)_3$.

Replacing $^{15}N$ by $^{14}N$ adds a further set of bands (see Supplementary Fig. S8). $^{15}N$ has I = 1/2 and no quadrupole moment, so it offers no $\Delta m_I = 2$ transition and cannot meet the cancellation condition, whereas $^{14}N$ has I = 1 and does. For the N-site defects the first coordination shell is boron, so these bands come from next-nearest-neighbor (NNN) nitrogen nuclei. They appear in the $^{11}B^{14}N$ bath near 4450 G for $C_N$, 5750 G for $C_2C_N$, and 9050 G for $C_NC_B^+$-DAP2. The mechanism is taken up below with the B-site defects.

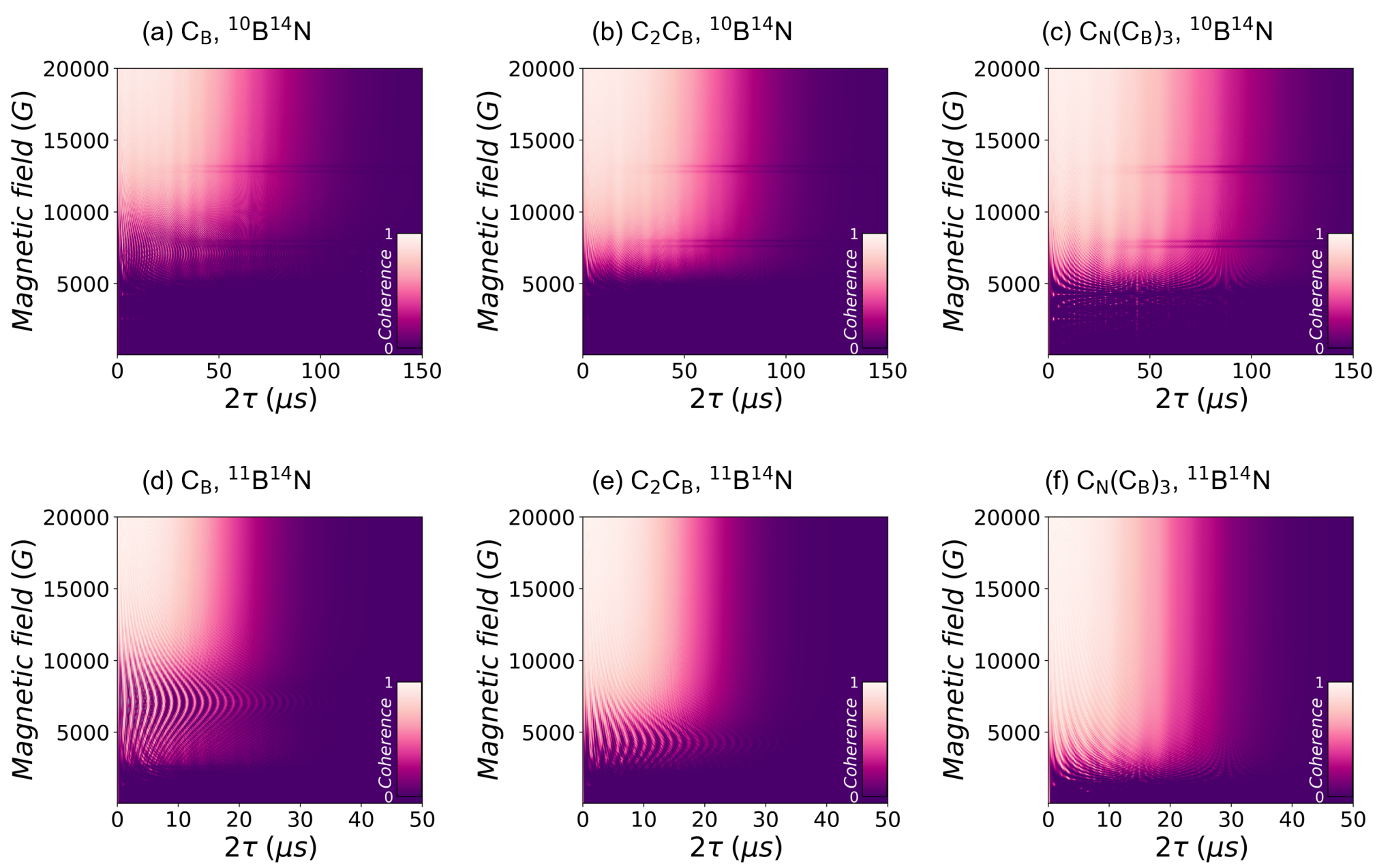

**Figure 4. Hahn-echo coherence maps of the three B-site defects in $^{14}N$-containing baths.** (a–c) $C_B$, $C_2C_B$, and $C_N(C_B)_3$ in the $^{10}B^{14}N$ bath; (d–f) the same defects in the $^{11}B^{14}N$ bath. Each map shows the coherence amplitude (color scale, 0 to 1) as a function of free-evolution time 2τ and magnetic field $B_0$. The triplet defect $C_N(C_B)_3$ is shown in the {$m_S$ = -1, $m_S$ = 0} qubit manifold.

The B-site defects provide a complementary test of the same mechanism, with the roles of boron and nitrogen neighbors reversed. Figure 4 shows their coherence maps in the $^{14}N$-containing baths ($^{10}B^{14}N$, Fig. 4a–c; $^{11}B^{14}N$, Fig. 4d–f). In the $^{10}B^{14}N$ bath only $C_B$ shows a modulation band, near 7100 G; in the $^{11}B^{14}N$ bath $C_B$ keeps that band at the same field and $C_2C_B$ gains one near 4200 G, while $C_N(C_B)_3$ shows none in either bath.

These observations follow from the properties of the $^{14}N$ neighbor. With I = 1 there is only a single $\Delta m_I = 2$ transition, and its two sublevels carry the same quadrupole shift, so the quadrupole contribution drops out and the cancellation field is set by the hyperfine coupling alone. It falls near 7150 G for $C_B$ and near 4400 G for $C_2C_B$; the latter lies below the 5000 G TB of the $^{10}B$ baths and is resolved only once the TB drops to about 2000 G in the $^{11}B^{14}N$ bath. Because the boron isotope does not enter the condition, $C_B$ modulates at the same field in both the $^{10}B^{14}N$ and $^{11}B^{14}N$ baths. Replacing $^{14}N$ by $^{15}N$ removes both bands altogether (Supplementary Fig. S1), confirming their $^{14}N$ origin. $C_N(C_B)_3$ shows no band because the sublevels of its nearest nitrogen never reach an anti-crossing at any field.

**Strong decoherence from heteronuclear pair resonances.** The third feature of Fig. 2 is a broad decoherence band centered near 15,000 G. The $T_2$ suppression appears at the same field for all four defects in the $^{10}B^{15}N$ bath (Fig. 2a–d) but is absent in the $^{11}B^{15}N$ bath (Fig. 2e–h), suggesting an origin in the bath rather than in the defect. To confirm this, we computed the decoherence of the $V_B^-$ spin in the $^{10}B^{15}N$ bath under the same conditions and find the same coherence reduction at the same field (see Supplementary Fig. S9). The decoherence band is therefore a bath property, controlled by the boron isotope.

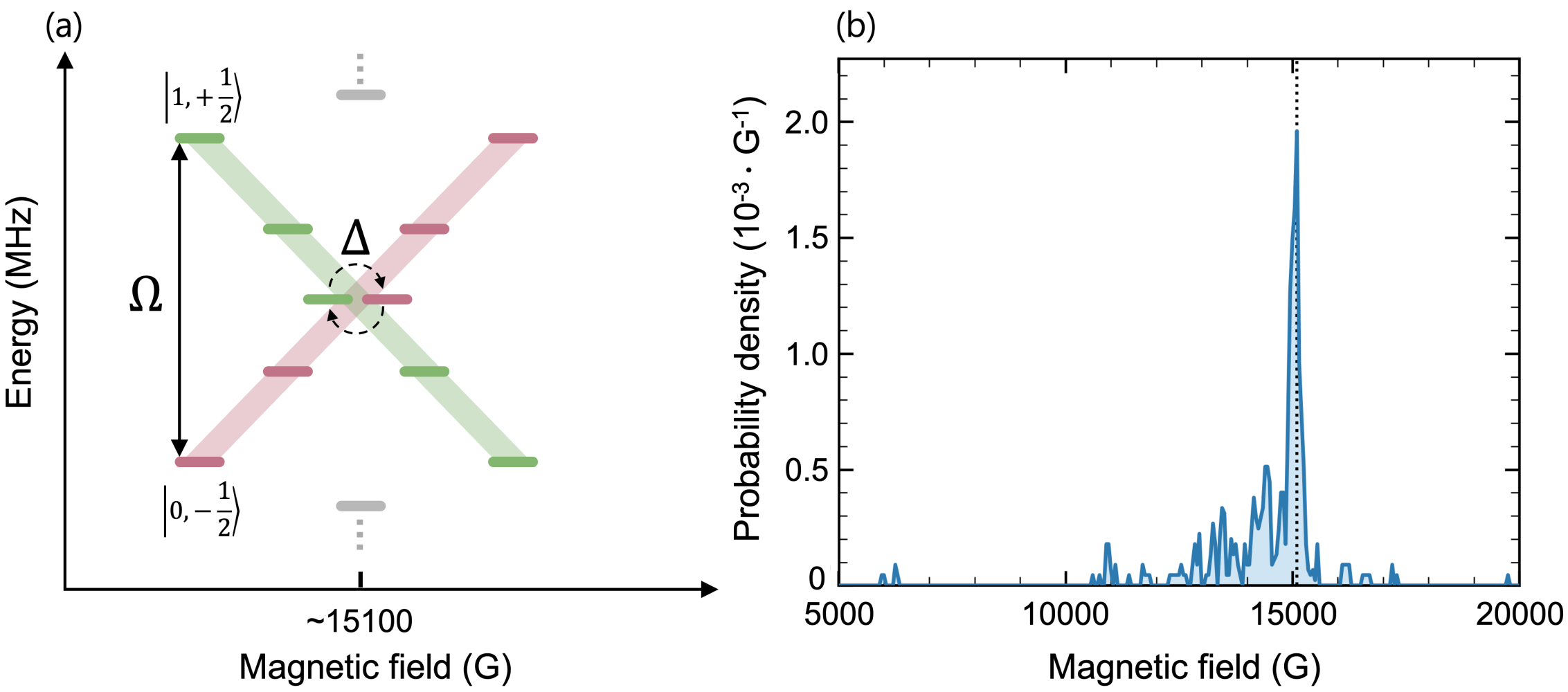

**Figure 5.** (a) Schematic energy levels (not to scale) of the two product states of a $^{10}$B (I = 3) – $^{15}$N (I = 1/2) pair, $|m_I(^{10}B) = 1, m_I(^{15}N) = +1/2\rangle$ (green) and $|0, -1/2\rangle$ (red), as a function of magnetic field. Their energy separation Ω vanishes near 15,100 G, where the pair dipolar interaction Δ couples them. Gray levels lie outside this two-level subspace. (b) Normalized distribution (in units of $10^{-3}$ $G^{-1}$) of $^{10}$B–$^{15}$N pairs satisfying $|\Delta/\Omega| \geq 1$ for the transition between $|1, +1/2\rangle$ and $|0, -1/2\rangle$, as a function of magnetic field, evaluated for every pair in the bath on a 50 G grid and normalized to unit area over the range shown, which lies above the TB. The dotted line marks the peak at 15,100 G.

As discussed above, decoherence above the TB is driven by dipolar transitions within pairs of bath nuclear spins. We find that this is also the case for the decoherence band near 15,100 G, but here the band originates from a resonance between heteronuclear $^{10}$B–$^{15}$N spin pairs. In particular, the two-state manifold consisting of $|m_I(^{10}B) = 1, m_I(^{15}N) = +1/2\rangle$ and $|0, -1/2\rangle$ undergoes rapid double-quantum transitions near 15,100 G, where the energy gap Ω between the two states becomes smaller than their dipolar coupling matrix element Δ (Fig. 5a). This is possible because the gyromagnetic ratios of $^{10}$B and $^{15}$N are nearly equal in magnitude but opposite in sign (4.57 and −4.32 MHz/T), so that the Zeeman contributions to Ω nearly cancel, leaving a residual field dependence of 26 Hz/G. The two states therefore approach degeneracy only near 15,100 G, where Δ drives the double-quantum transition. We also find that the $^{10}$B quadrupole interaction detunes the neighboring sublevel pairs, restricting the resonance to this pair of states.

To quantify these double-quantum transitions across the bath, Fig. 5b shows the distribution of $^{10}$B–$^{15}$N pairs satisfying $|\Delta/\Omega| \geq 1$ for this transition, evaluated as a function of magnetic field for all pairs included in the calculation. The distribution peaks at 15,100 G, coinciding with the decoherence band in the coherence maps. Its width of a few hundred gauss reflects the variation in local environments among pairs, which shifts the field at which each pair reaches degeneracy. In the $^{11}B^{15}N$ bath, the two gyromagnetic ratios differ substantially in magnitude, and the degeneracy condition falls below the TB, where dephasing from individual bath spins dominates; accordingly, no similar decoherence band appears. Because this band is determined by the isotopic composition of the bath rather than by the defect structure, it does not serve as a defect fingerprint; rather, it identifies which isotope combinations suffer coherence loss at high field.

**Magnetic-field dependence of $T_2$ below the TB.** In the low-field regime, we examine how $T_2$ varies with magnetic field. Figure 6a, b shows $T_2$ as a function of $B_0$ from 100 to 1000 G for all seven defects in the $^{10}B^{15}N$ and $^{11}B^{15}N$ baths, respectively. The two defect groups behave differently: N-site defects exhibit a pronounced increase in $T_2$ between 100 and 500 G, whereas B-site defects show only a weak field dependence. For $C_N$, $T_2$ increases 1.5-fold from 228 ns at 100 G to 333 ns at 500 G in the $^{10}B^{15}N$ bath, and 1.6-fold from 126 to 206 ns in the $^{11}B^{15}N$ bath. In contrast, $T_2$ of the B-site defects remain within 330–392 ns over the same field range in the $^{10}B^{15}N$ bath, and varies only weakly in the $^{11}B^{15}N$ bath. Above about 500 G the field dependence flattens for all seven defects, and the two groups merge into a common range of 308–392 ns at 1000 G. Based on separate analysis, we find that this contrasting behavior originates from the hyperfine-mediated interaction between nuclear spins [60], whose strength scales inversely with the electron spin splitting and thus grows at low field. The NN boron nuclei of the N-site defects, with their large nuclear spins, open many spin-pair transitions through this interaction and enhance decoherence, whereas the NN $^{15}N$ ($I = 1/2$) of the B-site defects supports only a single-pair transition, which the Hahn echo refocuses. The field dependence of $T_2$ below the TB therefore constitutes a group-level fingerprint, set by whether boron or nitrogen nuclei surround the defect.

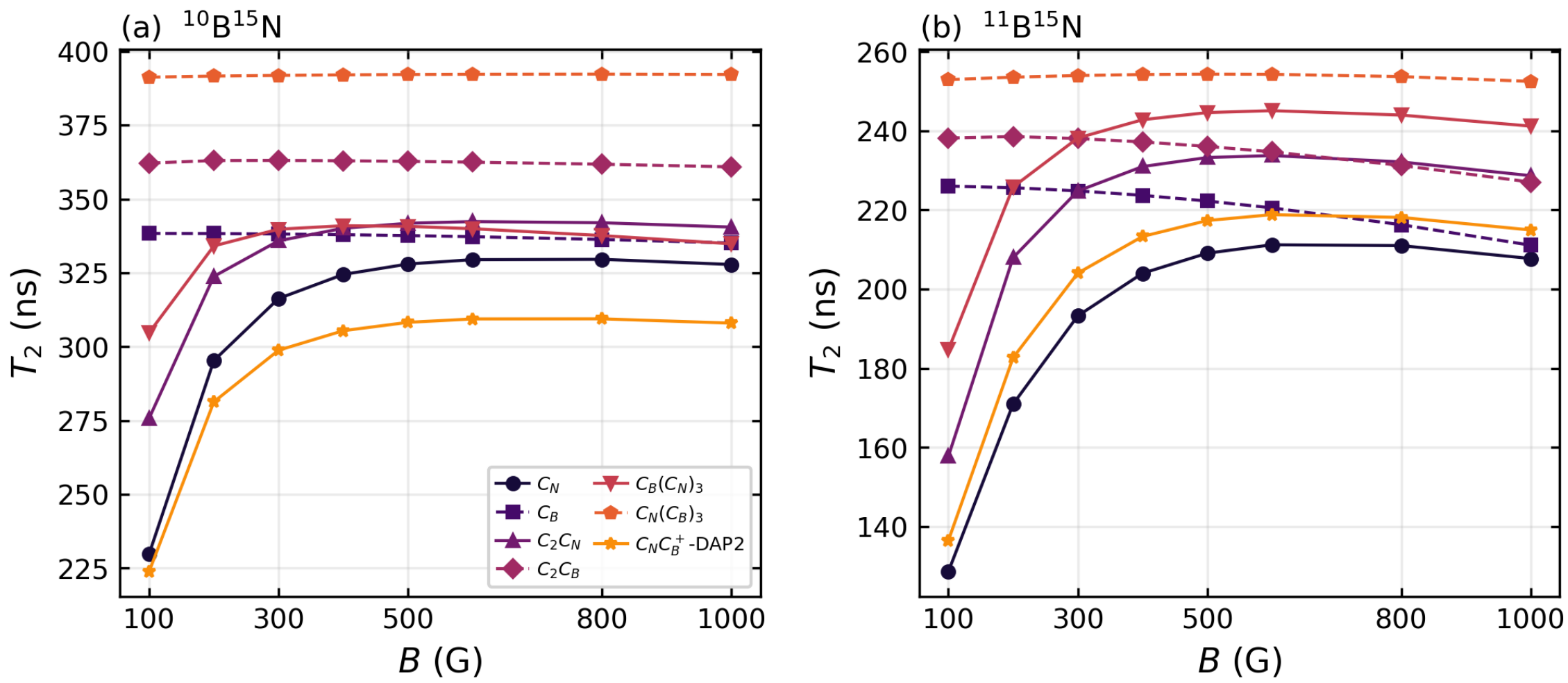

**Figure 6. Magnetic-field dependence of $T_2$ below the TB.** $T_2$ as a function of magnetic field for all seven defects in the $^{10}B^{15}N$ (a) and $^{11}B^{15}N$ (b) baths, computed at 100, 200, 300, 400, 500, 600, 800, and 1000 G. N-site defects show a pronounced field dependence below about 500 G, whereas B-site defects remain nearly field independent.

**CPMG coherence dynamics.** Finally, we examine whether dynamical decoupling can extend the coherence time of the carbon defects. We computed coherence functions under CPMG sequences with $N_\pi = 1$ (Hahn echo), 2, 4, 6, 10, 16, and 32 refocusing pulses for all seven defects in the four isotope baths at $B_0 = 500$ G. Figure 7a shows the results for $C_N$ in the $^{10}B^{15}N$ bath, where each increase in $N_\pi$ shifts the coherence decay to longer free-evolution times. Figure 7b–e plots $T_2$ as a function of $N_\pi$ on a log–log scale for all seven defects in the four baths. Although the curves are not perfectly linear, fitting each with the power law $T_2 \propto N_\pi^\gamma$ yields $\gamma = 0.80$–$0.90$ in the $^{10}B^{15}N$ bath. CPMG decoupling thus extends $T_2$ for all seven defects, but the narrow range of $\gamma$ means that, in contrast to the Hahn-echo fingerprints of the preceding sections, the scaling exponent offers little power to distinguish one defect from another.

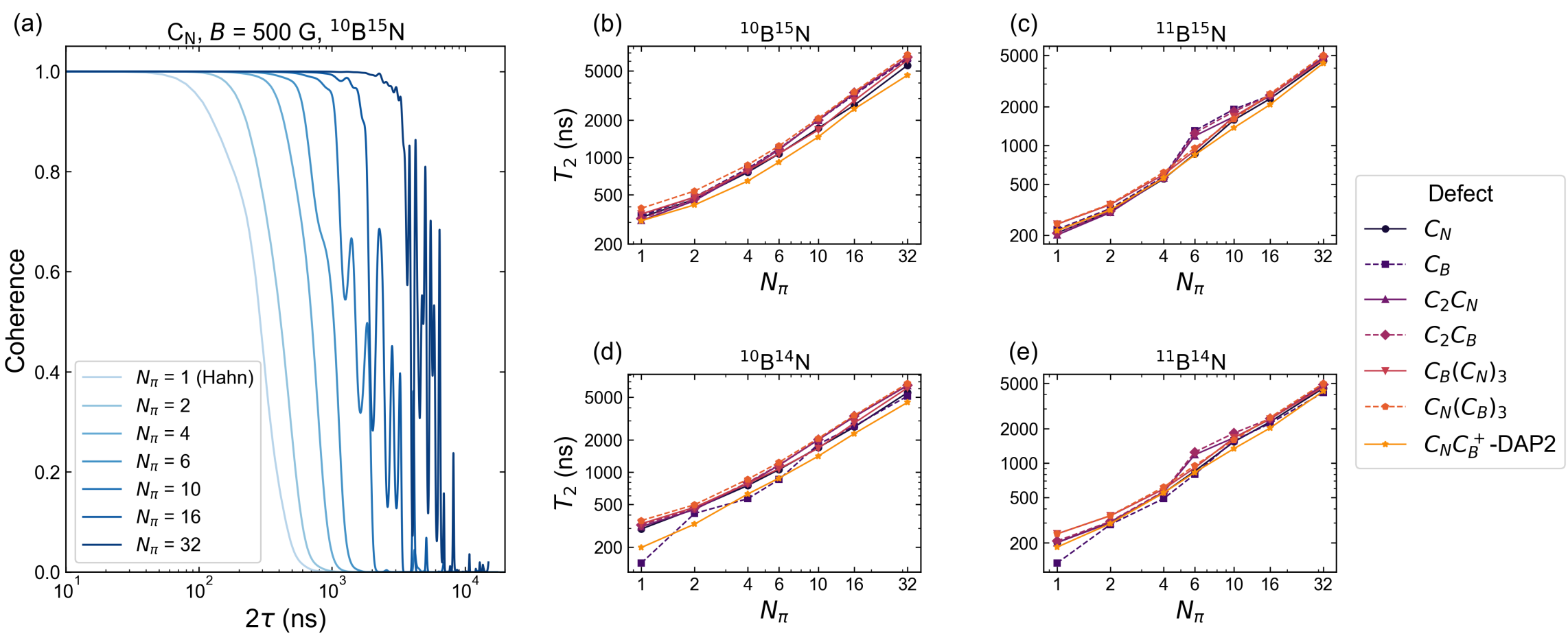

**Figure 7.** CPMG dynamical decoupling coherence dynamics at $B_0$ = 500 G. (a) Coherence functions of $C_N$ in the $^{10}B^{15}N$ bath under CPMG pulse sequences with $N_\pi$ = 1 (Hahn echo), 2, 4, 6, 10, 16, and 32. (b)–(e) $T_2$ scaling as a function of $N_\pi$ for all seven defects in the (b) $^{10}B^{15}N$, (c) $^{11}B^{15}N$, (d) $^{10}B^{14}N$, and (e) $^{11}B^{14}N$ baths, plotted on a log-log scale.

## Conclusions

In summary, we have investigated the electron-spin decoherence of seven carbon-related defects in h-BN from first principles, combining hybrid-functional DFT with cluster-correlation-expansion simulations across magnetic field and four isotope-engineered nuclear-spin baths. We find that each defect imprints its atomic structure on its coherence through two signatures. Above the transition boundary, ESEEM emerges at defect-specific magnetic fields at which the nearest-neighbor nuclear spins satisfy the cancellation condition; both the fields and the modulation frequencies follow from the computed hyperfine and quadrupole tensors alone, and they shift or vanish upon isotope substitution. Below the transition boundary, the magnetic-field dependence of $T_2$ separates the defects into N-site and B-site classes according to the sublattice occupied by carbon. Because these fingerprints can be computed in advance for any candidate structure, they establish a defect-identification framework complementary to optical spectroscopy and directly testable in currently available isotope-engineered h-BN samples.

Beyond defect identification, our results have direct implications for the coherence engineering of h-BN quantum sensors. CPMG dynamical decoupling extends $T_2$ for all seven defects, with scaling exponents of 0.80–0.90 in the $^{10}B^{15}N$ bath, although the exponent does not distinguish one defect from another. At high fields, double-quantum transitions of heteronuclear $^{10}B$–$^{15}N$ nuclear-spin pairs produce a decoherence band near 15,000 G, so isotope engineering of h-BN must consider which isotopes coexist in the bath, not only the spin properties of each species. These decoherence fingerprints, together with their predicted isotope and magnetic-field dependence, provide a practical basis for narrowing the structural assignment of carbon-related spin qubits in h-BN, a prerequisite for their rational control and deployment as nanoscale quantum sensors.

## Methods

**Density functional theory calculations.** All density functional theory (DFT) calculations were performed using the Vienna Ab-initio Simulation Package (VASP) [61] with the projector augmented-wave (PAW) method [62]. We employed the Heyd-Scuseria-Ernzerhof (HSE) screened hybrid exchange-correlation functional [63] with a mixing parameter of $\alpha = 0.32$ and a range-separation parameter of $\omega = 0.2$ Å$^{-1}$, which has been shown to accurately reproduce the band gap and defect properties of h-BN [36,54]. Van der Waals interactions between h-BN layers were included via the DFT-D3 method of Grimme [64]. The plane-wave kinetic energy cutoff was set to 600 eV. Bulk h-BN properties were computed using an $8 \times 8 \times 8$ k-point mesh for the primitive cell with AA' stacking.

Each carbon-related defect was modeled by introducing the corresponding substitutional carbon atom(s) into a $7 \times 4 \times 2$ orthogonal supercell of multilayer h-BN. Due to the large supercell size, the Brillouin zone was sampled at the Γ point only. All atomic positions were relaxed until the residual forces on each atom were less than 0.01 eV/Å. Supercell size convergence was verified by comparing the spin Hamiltonian parameters with those obtained from smaller supercells.

From the relaxed defect geometries, we computed the following spin Hamiltonian parameters for each of the seven defects, namely (i) the defect electronic structure and spin density distribution, (ii) hyperfine interaction tensors (both Fermi contact and anisotropic dipolar contributions) for all nuclear spins within the supercell, and (iii) nuclear quadrupole interaction tensors for nuclei with spin I ≥ 1. For the spin-triplet defects ($C_N(C_B)_3$ and $C_B(C_N)_3$), the zero-field splitting parameter D was taken from Benedek et al. [50], who obtained 820 MHz for $C_N(C_B)_3$ and 660 MHz for $C_B(C_N)_3$. The same computational approach was previously validated against experimental data for the $V_B^-$ defect in h-BN [54].

**Cluster correlation expansion calculations.** Electron spin decoherence was simulated using the cluster correlation expansion (CCE) method [55,56,58,65] at the second order (CCE-2) as implemented in the PyCCE code [57]. We used the secular approximation in which the $S_z$ component of the electron spin operator is retained while dropping the $S_x$ and $S_y$ components. The CCE calculations can additionally include the hyperfine-mediated interaction [60], in which virtual electron spin flips, driven by the off-diagonal electron spin matrix elements that are projected out in the secular approximation, generate an effective coupling between nuclear spin pairs via second-order perturbation theory. We note that the hyperfine-mediated interaction is essential for accurately describing the decoherence dynamics below about 500 G [54,60].

The spin Hamiltonian for the central qubit coupled to the nuclear spin bath comprises the electron Zeeman, zero-field splitting (for spin-triplet defects), hyperfine (secular and hyperfine-mediated), nuclear Zeeman, nuclear quadrupole, and nuclear dipole-dipole interactions. The full decomposition is given in Supplementary Information §S7.

The nuclear spin bath was constructed by including all boron and nitrogen nuclear spins within a bath radius of $r_{Bath}$ = 16 Å from the defect center, with pairwise nuclear dipolar couplings retained for separations no greater than $r_{Dip}$ = 6 Å. All production calculations use CCE-2. CCE-3 calculations at $B_0$ = 20000 G yield $T_2$ values about 20 % shorter than CCE-2 (Supplementary Figure S10), so CCE-2 slightly overestimates absolute $T_2$. Our defect-fingerprinting analysis relies on field-dependent patterns (cancellation-condition positions, $T_2$(B) trends, ESEEM modulation structure) rather than absolute $T_2$ values, and these features are well-converged at CCE-2. Convergence with respect to CCE cluster order, $r_{Bath}$, and $r_{Dip}$ is presented in Supplementary Figures S10–S12.

**Qubit basis states.** For spin-doublet (S = 1/2) defects ($C_N$, $C_B$, $C_2C_N$, $C_2C_B$, $C_NC_B^{+-}$DAP2), the qubit is defined as a spin-1/2 system using the $m_S = +1/2$ and $m_S = -1/2$ states. For spin-triplet (S = 1) defects ($C_N(C_B)_3$, $C_B(C_N)_3$), the qubit is defined as an effective two-level system using the $m_S$ = -1 and $m_S$ = 0 states. The {-1, 0} manifold was chosen because it is the manifold in which the cancellation condition between the hyperfine coupling and nuclear Zeeman interaction is met for $C_B(C_N)_3$, whose nearest $^{10}B$ carries a negative $A_{zz}$, and that condition is central to the defect fingerprinting analysis presented in this work. Additional analysis using the {+1, 0} manifold is given in Supplementary Figure S13.

## Acknowledgements

This study is supported by the National Research Foundation (NRF) of Korea grant funded by the Korean government (MSIT) (No. 2023R1A2C1006270 and No. RS-2025-25454922), by Creation of the Quantum Information Science R&D Ecosystem (Grant No. RS-2023-NR068116) through the NRF of Korea funded by the Korea government (MSIT). This work was supported by Institute of Information & communications Technology Planning & Evaluation (IITP) grant funded by the Korea government (MSIT) (RS-2025-25464252). This work was supported by the National Supercomputing Center with supercomputing resources including technical support (KSC-2025-CRE-0553).

## Conflict of Interest

The authors declare no conflict of interest.

## Table of Contents

Seven carbon-related spin defects in hexagonal boron nitride carry distinct magnetic-field-dependent coherence patterns. The spin echo is strongly modulated at a few specific fields, each set by strongly coupled nuclear spins, so the pattern moves or disappears when a host isotope is exchanged. These signatures form a structural fingerprint that complements optical identification.

# Supplementary Information for:

# Coherence-Based Identification of Carbon-Based Spin Qubits in Hexagonal Boron Nitride from First Principles

Hyeonsu Kim[1], Jaewook Lee[1], Huijin Park[1], and Hosung Seo[1,2]

[1]SKKU Advanced Institute of Nanotechnology, Sungkyunkwan University, Suwon, Gyeonggi 16419, Republic of Korea

[2]Department of Quantum Information Engineering, Sungkyunkwan University, Suwon, 16419, Republic of Korea

(correspondence to seo.hosung@skku.edu)

## Contents

## §S1. Hyperfine and quadrupole tensors

The hyperfine and quadrupole parameters of the symmetrically inequivalent atoms surrounding each carbon-related defect are tabulated below. All values are obtained from HSE06 ($\alpha = 0.32$) calculations on the relaxed geometries described in the Methods (main text). Tables S1–S7 report hyperfine tensors as their sorted principal eigenvalues ($A_1$, $A_2$, $A_3$ with $|A_1| \geq |A_2| \geq |A_3|$) in MHz, including the appropriate isotope gyromagnetic ratio. Tables S8–S14 report the electric-field-gradient (EFG) eigenvalues ($V_1$, $V_2$, $V_3$) in V/Å² surrounding each defect, sorted by magnitude in decreasing order ($|V_1| \geq |V_2| \geq |V_3|$). We note that the EFG tensor is traceless ($V_1 + V_2 + V_3 = 0$), and that EFG values are isotope-independent, so atom labels there omit the mass number, with nitrogen entries applying to both $^{14}N$ (quadrupole-active, I = 1) and $^{15}N$ (I = 1/2, no quadrupole).

**Atom-label convention.** Distances are measured from the spin-bearing carbon at the center of each defect, which is denoted $^{13}C$-center, or $^{13}C_X$-center for the cluster defects, where X identifies the sublattice site it occupies. The remaining atoms are numbered by increasing distance, B1, B2, … for boron and N1, N2, … for nitrogen.

**Table S1.** Hyperfine interaction parameters for the $C_N$ defect (S = 1/2) in h-BN. Values in parentheses are the corresponding eigenvalues from Auburger and Gali[1].

| Atom | Distance (Å) | $A_1$ (MHz) | $A_2$ (MHz) | $A_3$ (MHz) |
|---|---|---|---|---|
| $^{13}C$-center | 0.00 | 152.56 (156.52) | -19.45 (-19.21) | -19.45 (-19.21) |
| $^{10}B1$ | 1.51 | -8.02 (-8.21) | -5.93 (-6.01) | -5.45 (-5.36) |

**Table S2.** Hyperfine interaction parameters for the $C_B$ defect (S = 1/2) in h-BN. Values in parentheses are the corresponding eigenvalues from Auburger and Gali[1].

| Atom | Distance (Å) | $A_1$ (MHz) | $A_2$ (MHz) | $A_3$ (MHz) |
|---|---|---|---|---|
| $^{13}C$-center | 0.00 | 236.60 (231.29) | 14.36 (12.09) | 14.36 (12.09) |
| $^{15}N1$ | 1.40 | 12.51 (12.66) | 12.26 (12.57) | 6.18 (7.17) |

**Table S3.** Hyperfine interaction parameters for the $C_2C_N$ defect (S = 1/2) in h-BN. Values in parentheses are the corresponding eigenvalues from Auburger and Gali[1].

| Atom | Distance (Å) | $A_1$ (MHz) | $A_2$ (MHz) | $A_3$ (MHz) |
|---|---|---|---|---|
| $^{13}C_B$-center | 0.00 | -66.06 (-69.58) | -28.18 (-28.68) | -25.00 (-26.18) |
| $^{13}C_{N1}$ | 1.39 | 107.54 (110.17) | -7.11 (-6.64) | -5.89 (-5.78) |
| $^{15}$N1 | 1.40 | 1.98 (1.83) | -1.51 (-1.49) | -1.05 (-1.15) |
| $^{10}$B1 | 2.54 | -5.08 (-5.18) | -4.05 (-4.12) | -2.53 (-2.58) |

**Table S4.** Hyperfine interaction parameters for the $C_2C_B$ defect (S = 1/2) in h-BN. Values in parentheses are the corresponding eigenvalues from Auburger and Gali[1].

| Atom | Distance (Å) | $A_1$ (MHz) | $A_2$ (MHz) | $A_3$ (MHz) |
|---|---|---|---|---|
| $^{13}C_N$-center | 0.00 | -66.50 (-67.81) | -34.06 (-33.46) | -28.73 (-29.39) |
| $^{13}C_{B1}$ | 1.39 | 148.92 (150.73) | 12.65 (12.61) | 11.17 (11.50) |
| $^{10}$B1 | 1.52 | 1.82 (1.83) | 1.70 (1.76) | 0.66 (0.75) |
| $^{15}$N1 | 2.42 | 6.35 (6.56) | 5.99 (6.19) | -3.81 (-2.16) |

**Table S5.** Hyperfine interaction parameters for the $C_N(C_B)_3$ defect (S = 1) in h-BN. Values in parentheses are the corresponding eigenvalues from Benedek *et al.*[2].

| Atom | Distance (Å) | $A_1$ (MHz) | $A_2$ (MHz) | $A_3$ (MHz) |
|---|---|---|---|---|
| $^{13}C_N$-center | 0.00 | -57.38 (-57.60) | -33.14 (-33.40) | -33.14 (-33.40) |
| $^{13}C_{B1}$ | 1.41 | 94.68 (95.00) | 8.03 (8.80) | 7.70 (8.60) |
| $^{15}$N1 | 2.45 | 4.69 (4.63) | 4.33 (4.49) | -0.40 (-0.42) |

**Table S6.** Hyperfine interaction parameters for the $C_B(C_N)_3$ defect (S = 1) in h-BN. Values in parentheses are the corresponding eigenvalues from Benedek *et al.*[2].

| Atom | Distance (Å) | $A_1$ (MHz) | $A_2$ (MHz) | $A_3$ (MHz) |
|---|---|---|---|---|
| $^{13}C_B$-center | 0.00 | -47.57 (-48.00) | -22.55 (-22.90) | -22.55 (-22.90) |
| $^{13}C_{N1}$ | 1.40 | 68.18 (68.90) | -3.26 (-2.60) | -2.90 (-2.60) |
| $^{10}$B1 | 2.54 | -3.52 (-3.55) | -2.81 (-2.85) | -2.23 (-2.28) |

**Table S7.** Hyperfine interaction parameters for the $C_NC_B^+$–DAP2 defect (S = 1/2) in h-BN. Values in parentheses are the corresponding eigenvalues from Auburger and Gali[1].

| Atom | Distance (Å) | $A_1$ (MHz) | $A_2$ (MHz) | $A_3$ (MHz) |
|---|---|---|---|---|
| $^{13}$C-center | 0.00 | 148.59 (150.24) | -18.57 (-19.42) | -18.47 (-19.34) |
| $^{10}$B1 | 1.49 | -7.83 (-8.04) | -5.72 (-5.79) | -4.69 (-4.65) |
| $^{10}$B2 | 1.51 | -7.54 (-7.67) | -5.87 (-5.85) | -5.63 (-5.64) |

**Table S8.** Eigenvalues of the EFG tensor (V/Å²) for the $C_N$ defect in h-BN. Values in parentheses are the corresponding eigenvalues from Auburger and Gali[1].

| Atom | Distance (Å) | $V_1$ (V/Å²) | $V_2$ (V/Å²) | $V_3$ (V/Å²) |
|---|---|---|---|---|
| B1 | 1.51 | 42.83 (43.16) | -26.32 (-26.93) | -16.51 (-16.23) |

**Table S9.** Eigenvalues of the EFG tensor (V/Å²) for the $C_B$ defect in h-BN. Values in parentheses are the corresponding eigenvalues from Auburger and Gali[1].

| Atom | Distance (Å) | $V_1$ (V/Å²) | $V_2$ (V/Å²) | $V_3$ (V/Å²) |
|---|---|---|---|---|
| N1 | 1.40 | 45.51 (46.21) | -27.34 (-30.84) | -18.18 (-15.37) |

**Table S10.** Eigenvalues of the EFG tensor (V/Å²) for the $C_2C_N$ defect in h-BN. Values in parentheses are the corresponding eigenvalues from Auburger and Gali[1].

| Atom | Distance (Å) | $V_1$ (V/Å²) | $V_2$ (V/Å²) | $V_3$ (V/Å²) |
|---|---|---|---|---|
| N1 | 1.40 | 36.51 (38.52) | -21.80 (-25.85) | -14.71 (-12.67) |
| B1 | 2.54 | 42.10 (42.49) | -25.06 (-25.59) | -17.04 (-16.89) |

**Table S11.** Eigenvalues of the EFG tensor (V/Å²) for the $C_2C_B$ defect in h-BN. Values in parentheses are the corresponding eigenvalues from Auburger and Gali[1].

| Atom | Distance (Å) | $V_1$ (V/Å²) | $V_2$ (V/Å²) | $V_3$ (V/Å²) |
|---|---|---|---|---|
| B1 | 1.52 | 40.90 (41.33) | -24.59 (-25.21) | -16.31 (-16.12) |
| N1 | 2.42 | 40.97 (42.67) | -26.29 (-30.03) | -14.68 (-12.64) |

**Table S12.** Eigenvalues of the EFG tensor (V/Å²) for the $C_N(C_B)_3$ defect in h-BN.

| Atom | Distance (Å) | $V_1$ (V/Å²) | $V_2$ (V/Å²) | $V_3$ (V/Å²) |
|---|---|---|---|---|
| N1 | 2.45 | 44.15 | -28.03 | -16.11 |

**Table S13.** Eigenvalues of the EFG tensor (V/Å²) for the $C_B(C_N)_3$ defect in h-BN.

| Atom | Distance (Å) | $V_1$ (V/Å²) | $V_2$ (V/Å²) | $V_3$ (V/Å²) |
|---|---|---|---|---|
| B1 | 2.54 | 42.56 | -25.70 | -16.86 |

**Table S14.** Eigenvalues of the EFG tensor (V/Å²) for the $C_NC_B^+$–DAP2 defect in h-BN. Values in parentheses are the corresponding eigenvalues from Auburger and Gali[1].

| Atom | Distance (Å) | $V_1$ (V/Å²) | $V_2$ (V/Å²) | $V_3$ (V/Å²) |
|---|---|---|---|---|
| B1 | 1.49 | 41.30 (41.67) | -30.15 (-31.01) | -11.15 (-10.66) |
| B2 | 1.51 | 42.27 (42.70) | -23.97 (-24.69) | -18.30 (-18.01) |

## §S2. Coherence maps of the B-site defects in the $^{10}B^{15}N$ and $^{11}B^{15}N$ baths

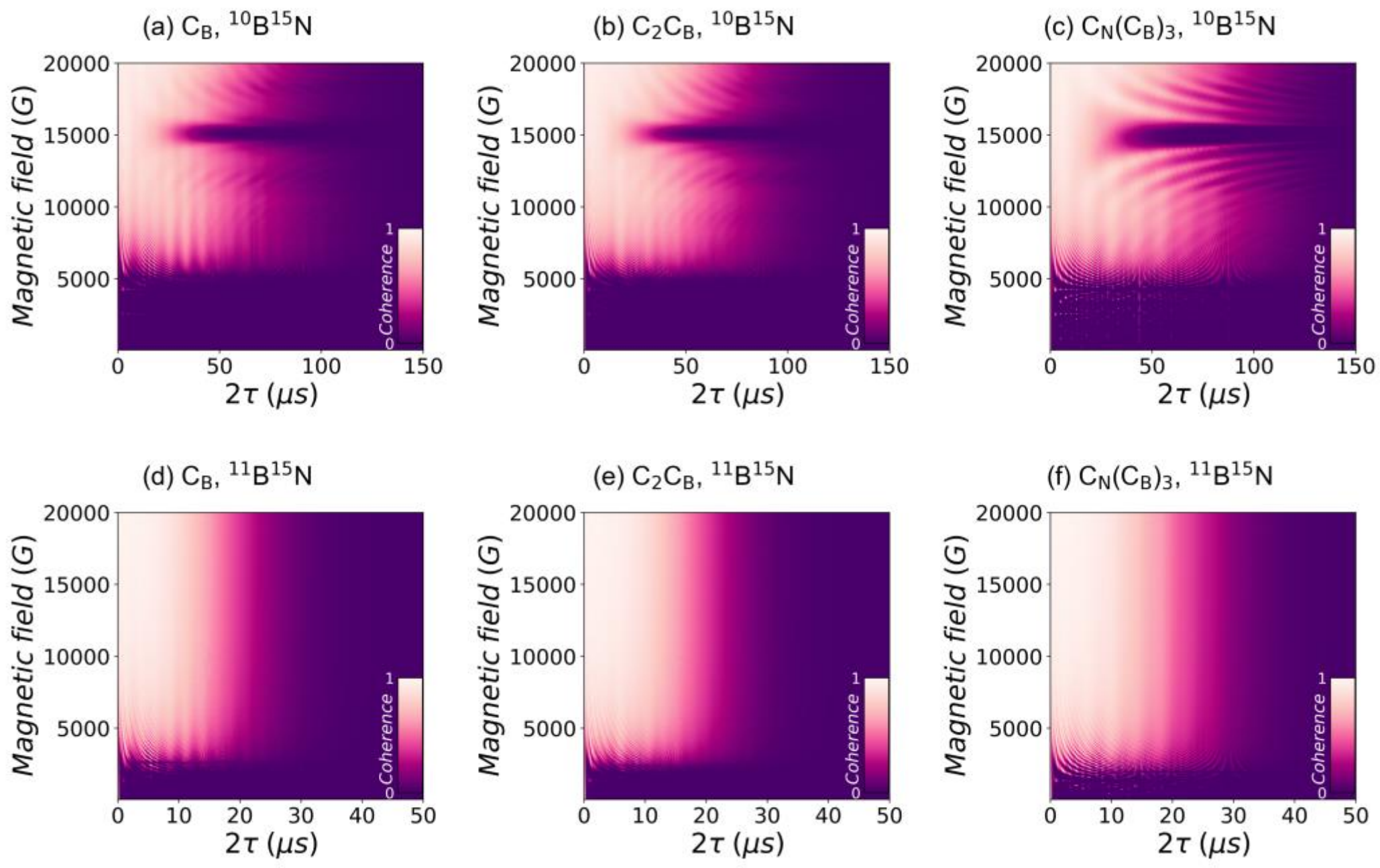

**Figure S1.** Hahn-echo coherence colormaps for the three B-site defects in $^{15}N$-containing baths.

## §S3. Pseudo-spin model of the cancellation-condition modulation

Near a cancellation field, the relevant pair of nuclear sublevels can be mapped onto an effective I=1/2 pseudo-spin coupled to the qubit. The corresponding Hahn-echo coherence takes the standard two-pulse ESEEM form [3, 4],

$$L(2\tau) = 1 - 2k\sin^2\left(\frac{\omega_+\tau}{2}\right)\sin^2\left(\frac{\omega_-\tau}{2}\right), \qquad \text{(S1)}$$

where $\omega_+$ and $\omega_-$ are the nuclear transition frequencies in the two qubit spin states and k is the modulation depth of Eq. (S3). Because the coherence is followed as a function of the total evolution time $2\tau$, a transition at $\omega$ modulates it at $\omega/2$, so the peaks in the Fourier spectrum of Figure 3(b) lie at $\omega_-/2$ and $\omega_+/2$.

The two frequencies follow from the pseudo-spin Hamiltonian,

$$\omega_\pm = \left[\left(m_S^\pm nA + n\,\omega_I + \Omega_Q\right)^2 + 4\left(m_S^\pm B + \Delta_Q\right)^2\right]^{1/2}, \qquad \text{(S2)}$$

with $m_S^\pm$ the qubit spin projection of each manifold, A the secular hyperfine constant, B the pseudo-secular hyperfine constant, $\omega_I$ the nuclear Zeeman frequency, which equals $-\gamma_n B_0$ and so carries the sign of the gyromagnetic ratio, $\Omega_Q$ and $\Delta_Q$ the quadrupole-induced diagonal splitting and off-diagonal coupling of the two nuclear sublevels, and n = 1 for a single-quantum transition or n = 2 for a double-quantum transition.

The modulation depth combines the ESEEM mixing of the two sublevels with their population in the maximally mixed nuclear state,

$$k = \frac{2}{2I+1}\left[\frac{2\left(\Delta_Q\, nA - B\left(n\,\omega_I + \Omega_Q\right)\right)}{\omega_+\omega_-}\right]^2, \qquad \text{(S3)}$$

where the factor 2/(2I + 1) is the population weight of the two sublevels of the selected transition and the bracketed term is the mixing parameter set by the spin Hamiltonian.

## §S4. Cancellation conditions for the seven defects

Cancellation fields were extracted from CCE-2 simulations by identifying local maxima in the modulation strength as a function of $B_0$, with FFT analysis used to validate each feature. The field uncertainty is ±50 G, set by the simulation grid.

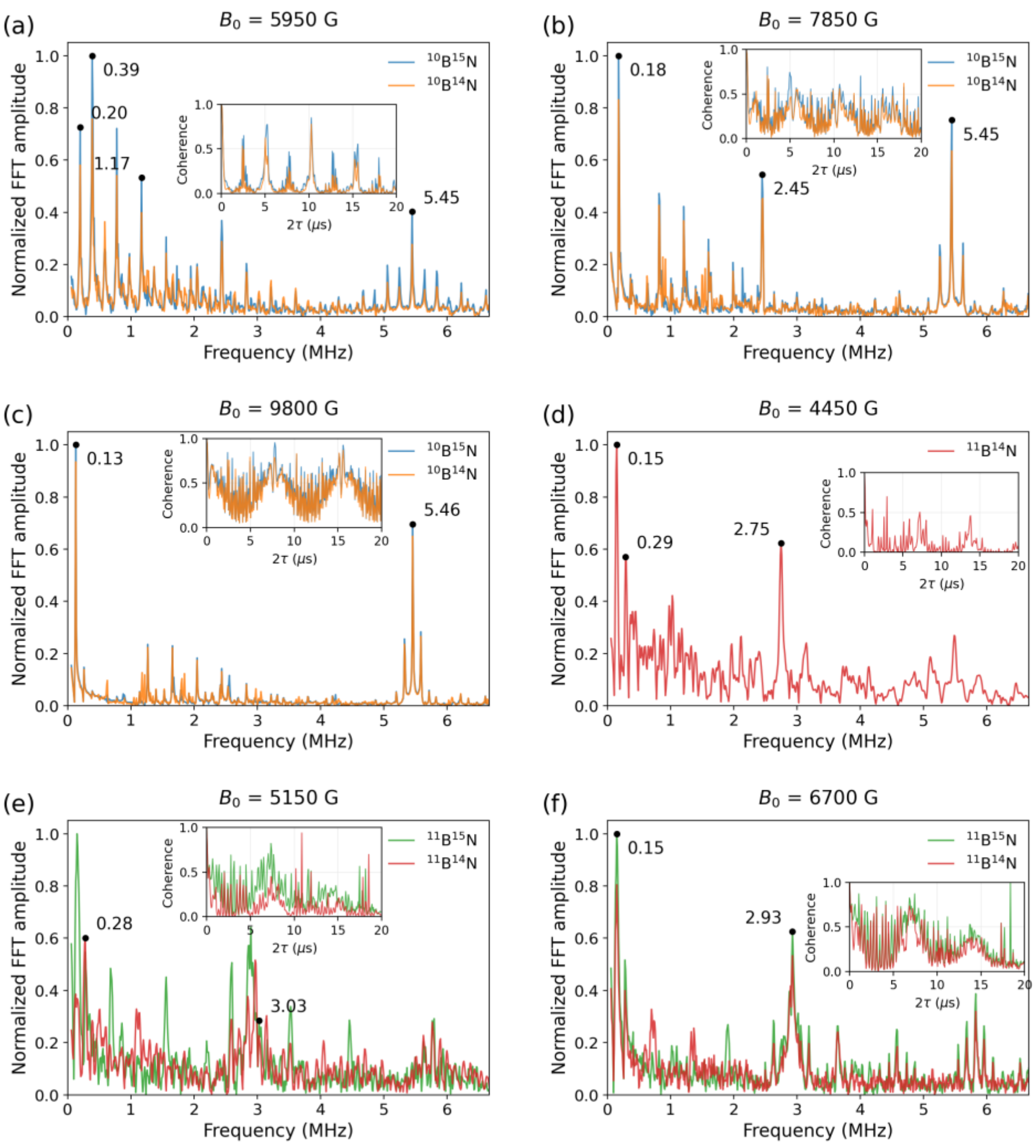

**Figure S2.** Cancellation-field FFT spectra for the $C_N$ defect. Each panel shows the FFT amplitude of the Hahn-echo coherence at the indicated magnetic field, with the corresponding coherence function over a 0-20 μs window shown in the inset. Results for isotope baths sharing the same cancellation field are overlaid. Black markers indicate the prominent FFT peaks.

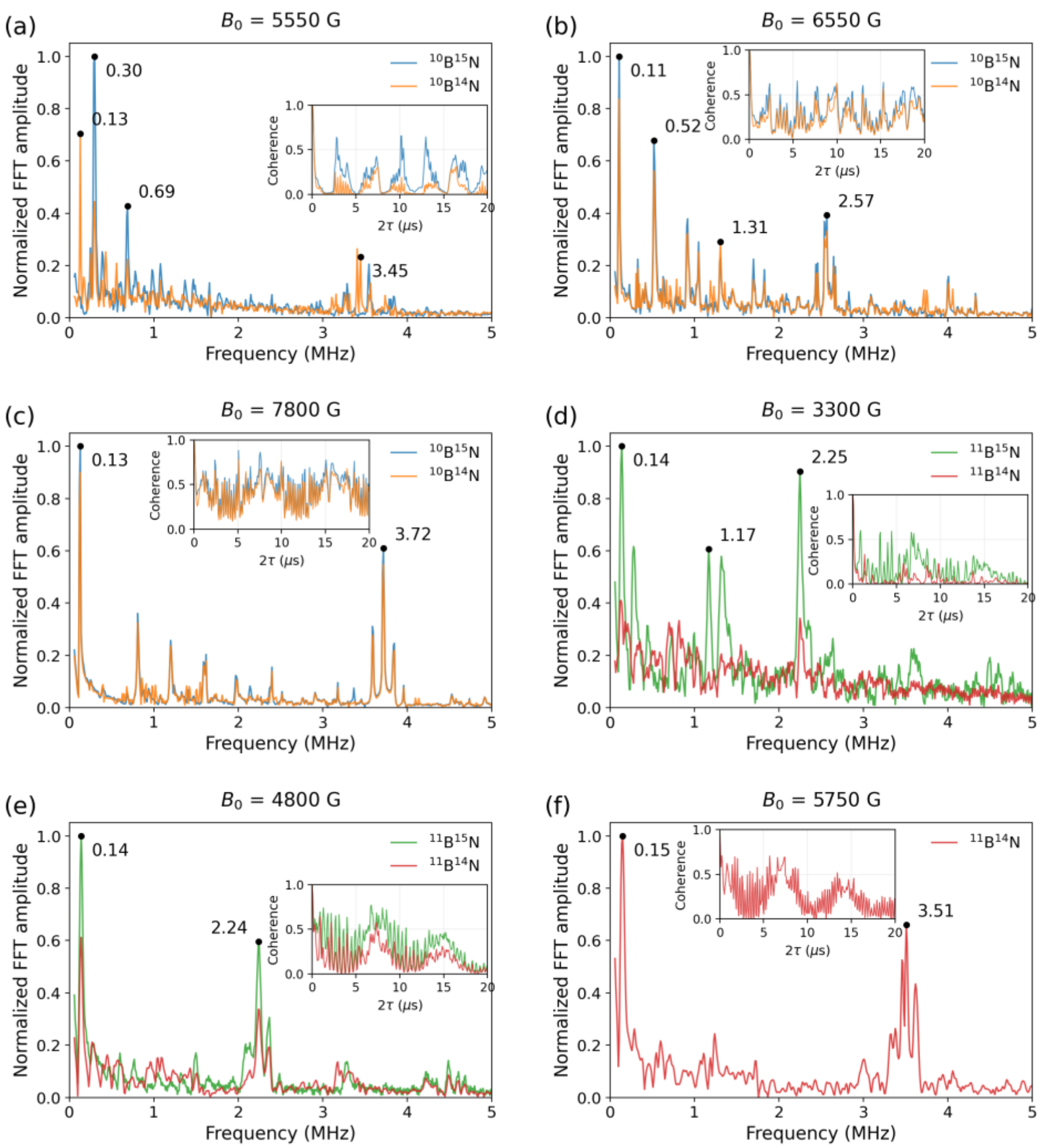

**Figure S3.** Cancellation-field FFT spectra for the $C_2C_N$ defect. Each panel shows the FFT amplitude of the Hahn-echo coherence at the indicated magnetic field, with the corresponding coherence function over a 0-20 μs window shown in the inset. Results for isotope baths sharing the same cancellation field are overlaid. Black markers indicate the prominent FFT peaks.

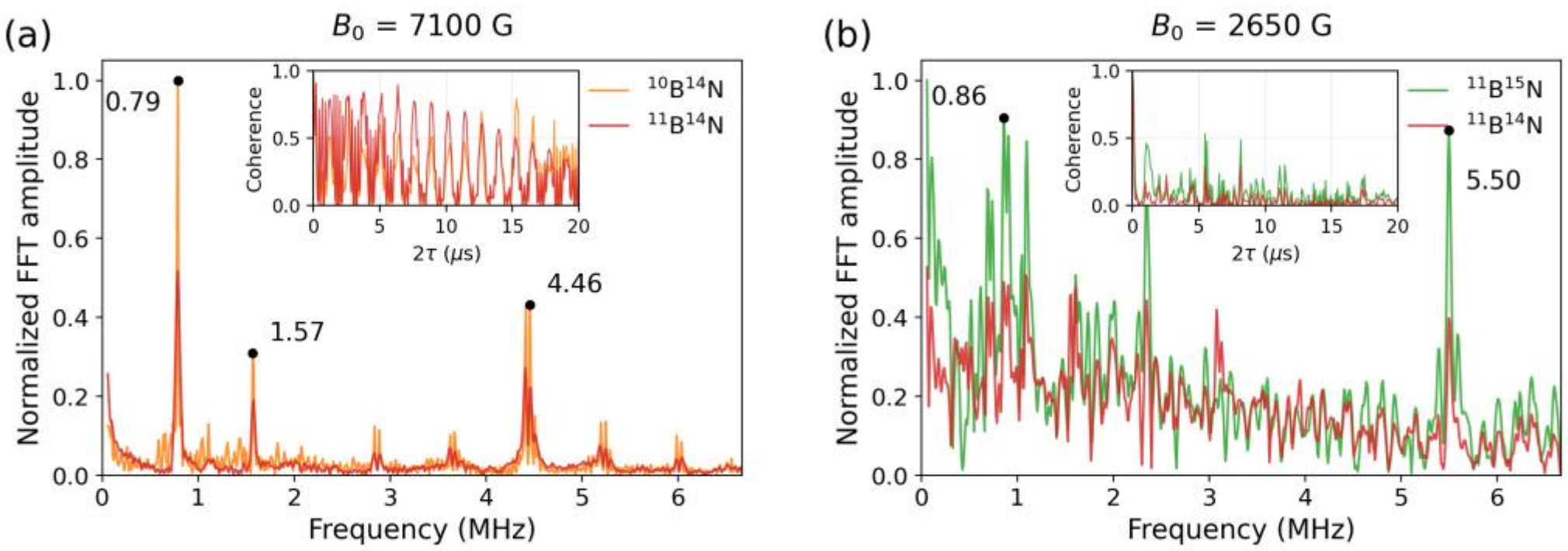

**Figure S4.** Cancellation-field FFT spectra for the $C_B$ defect. Each panel shows the FFT amplitude of the Hahn-echo coherence at the indicated magnetic field, with the corresponding coherence function over a 0-20 μs window shown in the inset. Results for isotope baths sharing the same cancellation field are overlaid. Black markers indicate the prominent FFT peaks.

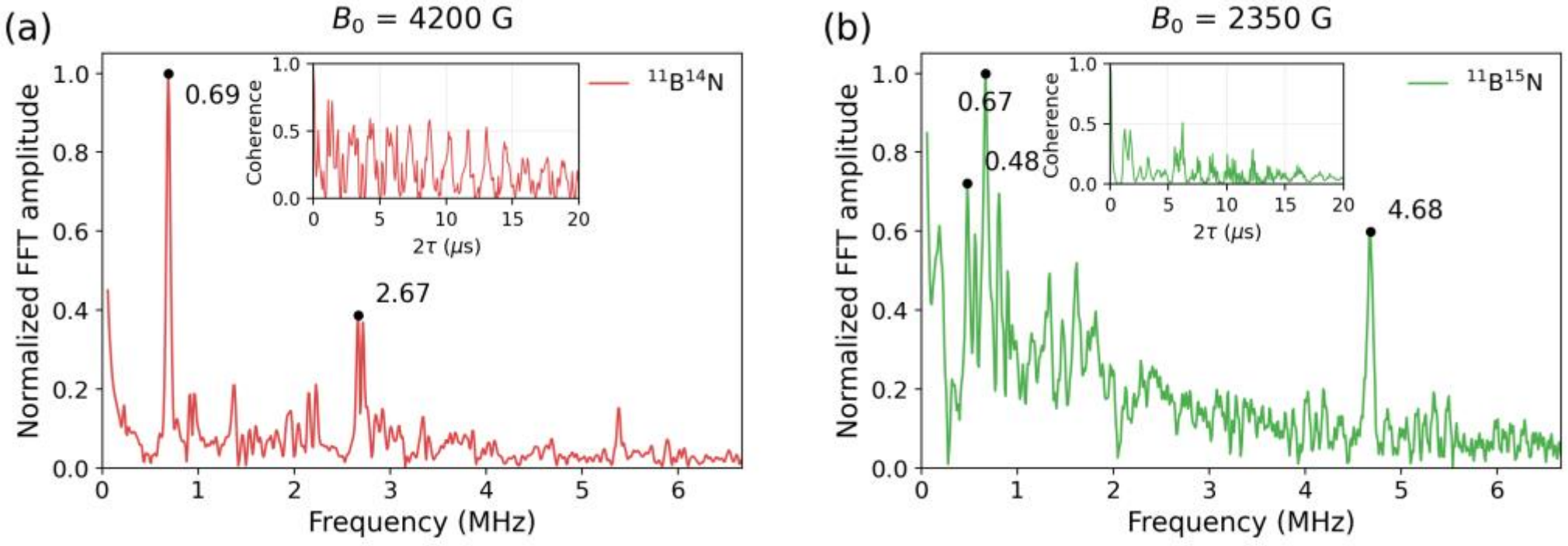

**Figure S5.** Cancellation-field FFT spectra for the $C2C_B$ defect. Each panel shows the FFT amplitude of the Hahn-echo coherence at the indicated magnetic field, with the corresponding coherence function over a 0-20 μs window shown in the inset. Results for isotope baths sharing the same cancellation field are overlaid. Black markers indicate the prominent FFT peaks.

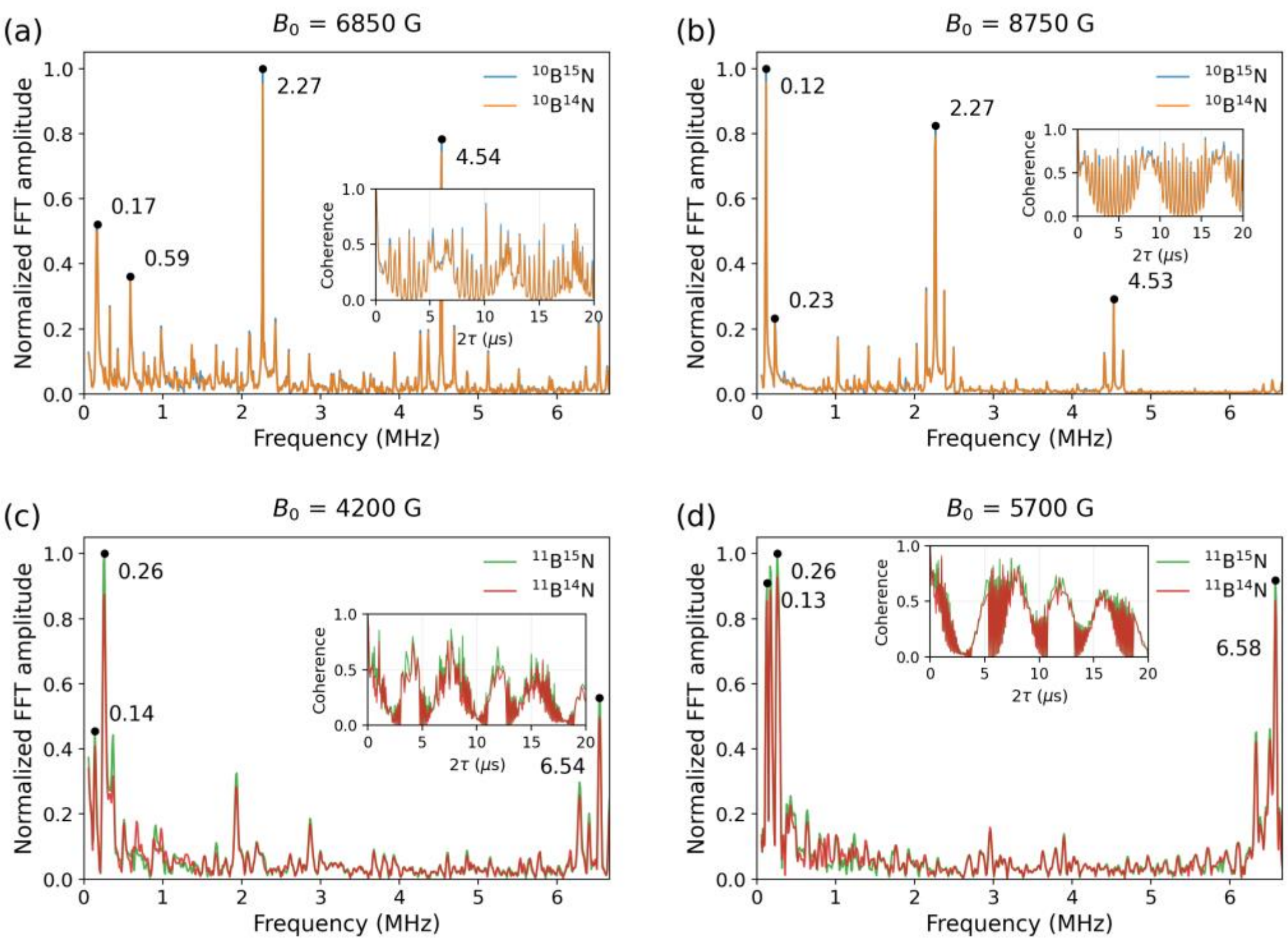

**Figure S6.** Cancellation-field FFT spectra for the $C_B(C_N)_3$ defect. Each panel shows the FFT amplitude of the Hahn-echo coherence at the indicated magnetic field, with the corresponding coherence function over a 0-20 μs window shown in the inset. Results for isotope baths sharing the same cancellation field are overlaid. Black markers indicate the prominent FFT peaks.

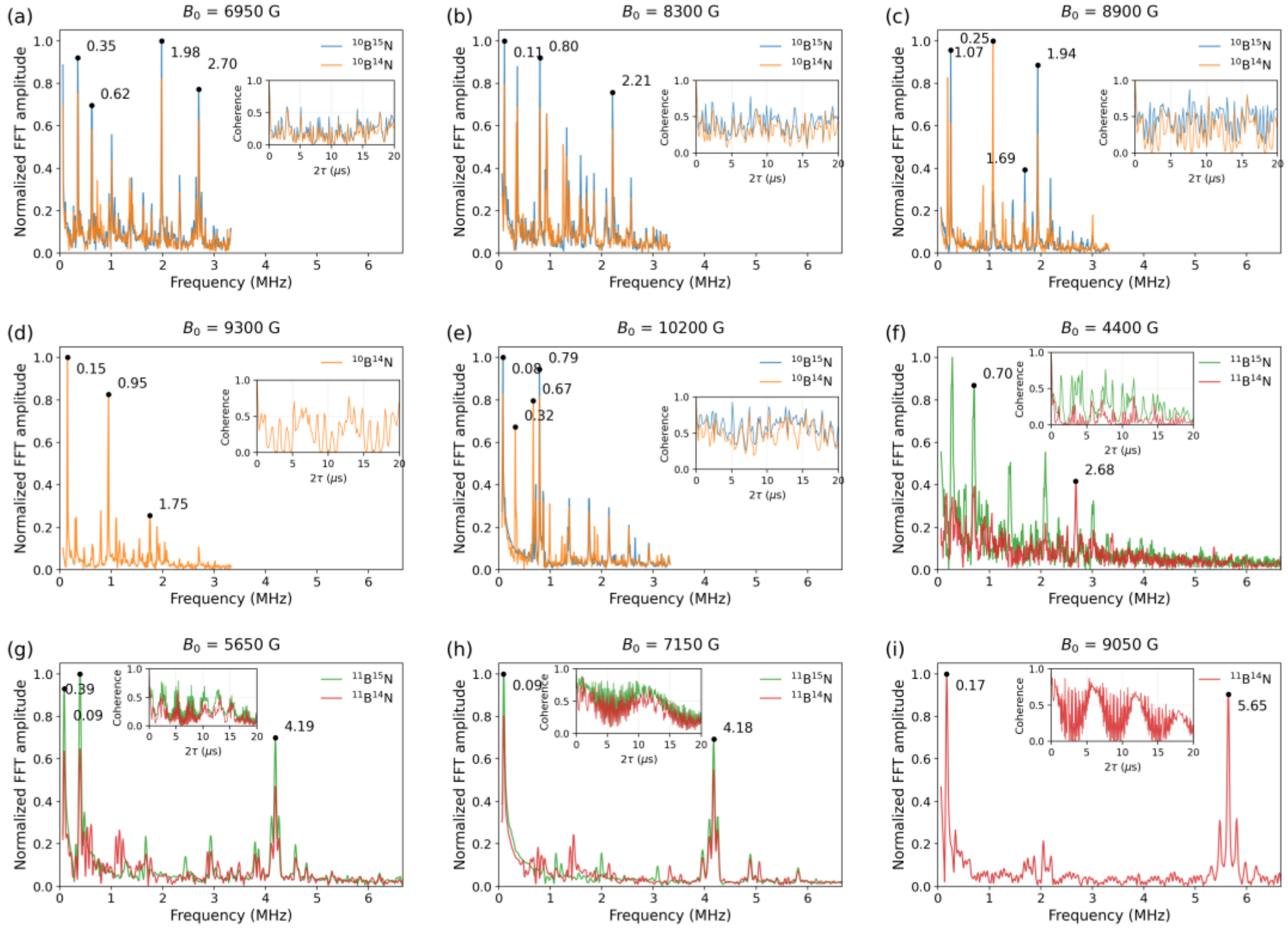

**Figure S7.** Cancellation-field FFT spectra for the $C_NC_B^+$-DAP2 defect. Each panel shows the FFT amplitude of the Hahn-echo coherence at the indicated magnetic field, with the corresponding coherence function over a 0-20 μs window shown in the inset. Results for isotope baths sharing the same cancellation field are overlaid. Black markers indicate the prominent FFT peaks.

## §S5. Coherence maps of the N-site defects in the $^{10}B^{14}N$ and $^{11}B^{14}N$ baths

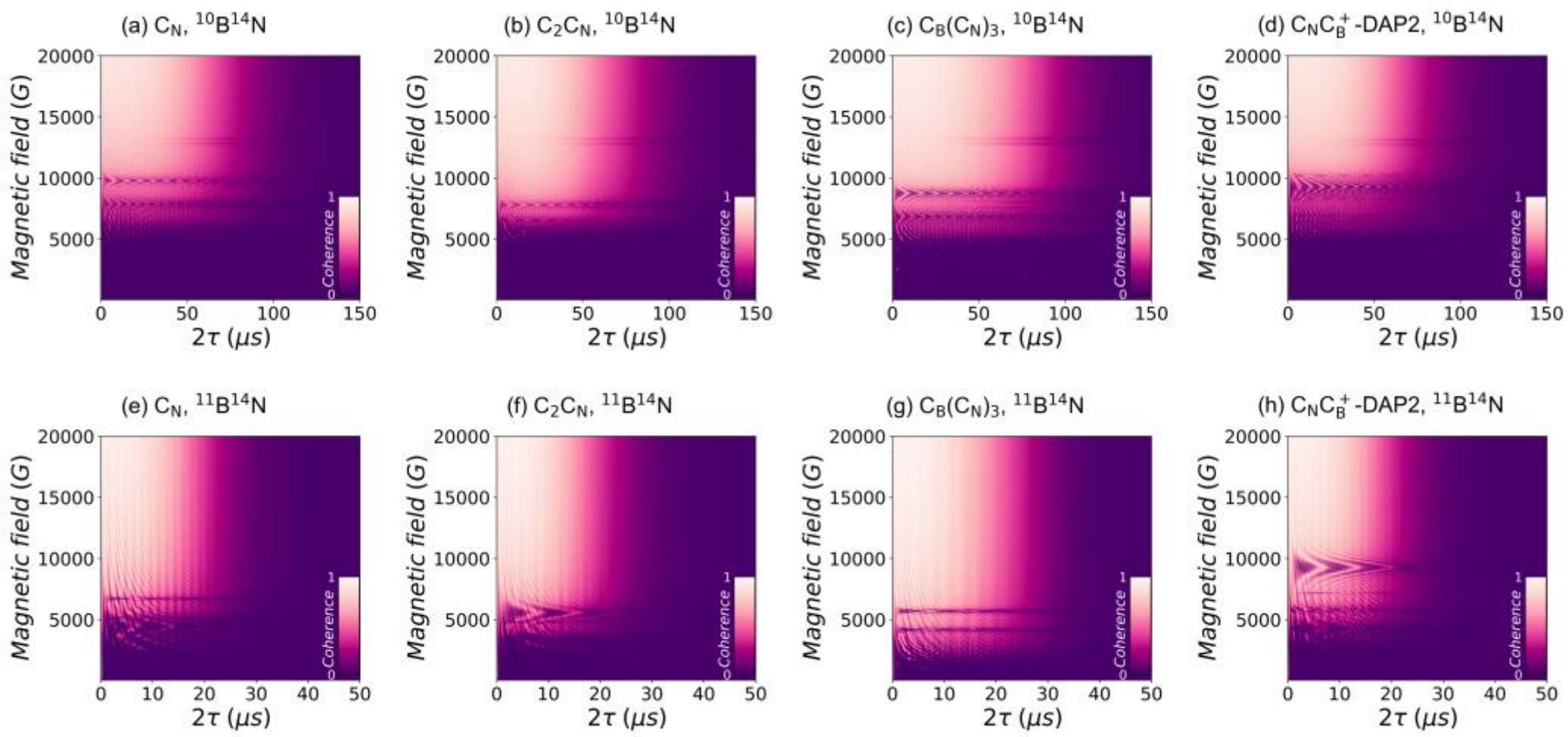

**Figure S8.** Hahn-echo coherence colormaps for the four N-site defects in $^{14}N$-containing baths.

## §S6. Coherence map of $V_B^-$ defect in h-$^{10}B^{15}N$ bath

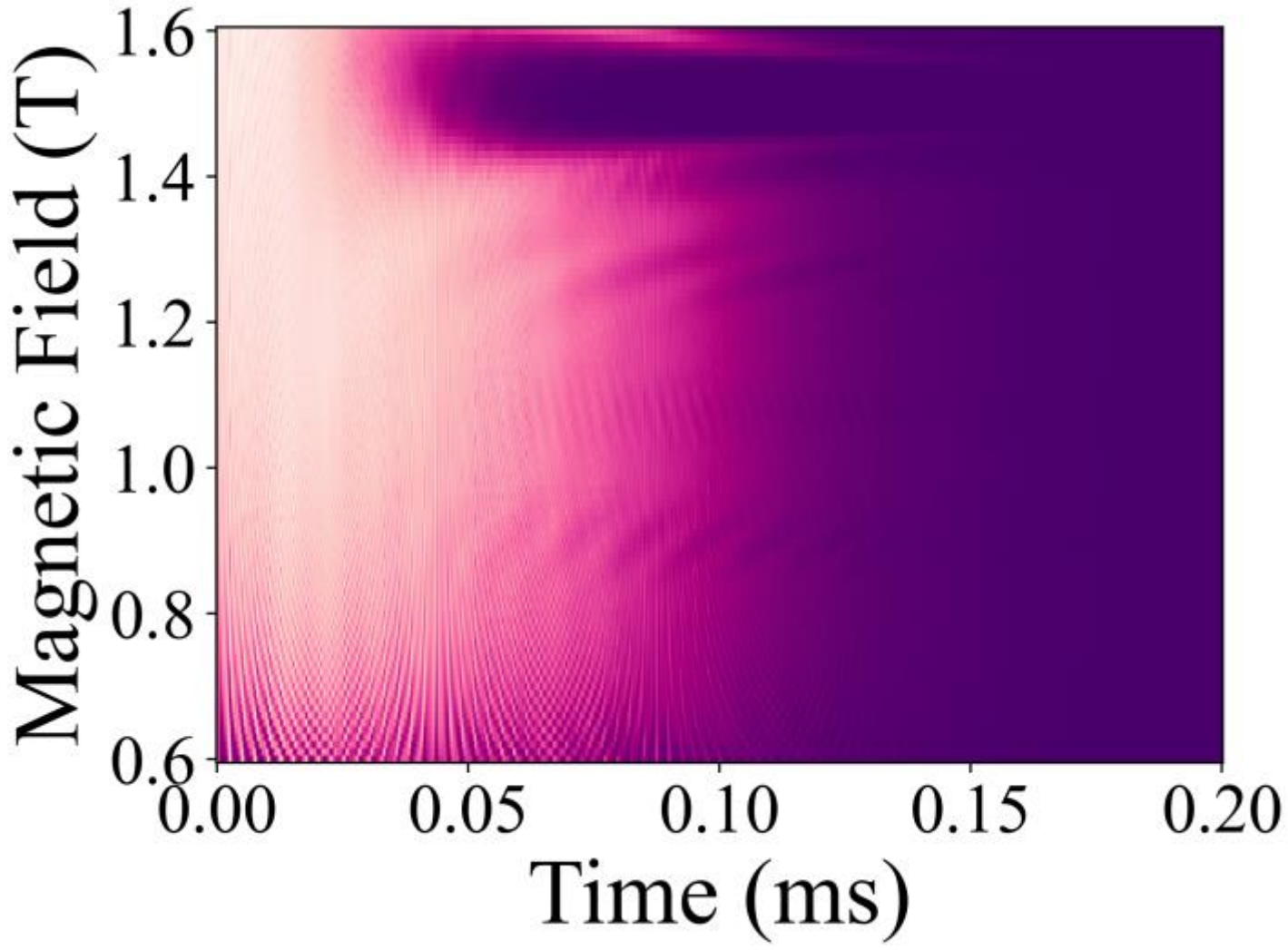

**Figure S9.** Hahn-echo coherence colormap of the $V_B^-$ defect in the $^{10}B^{15}N$ bath, computed with the same CCE-2 parameters as the carbon defects. Shown to demonstrate that the broad $T_2$ reduction near 15000 G is a bath-intrinsic property, not specific to carbon defects, since $V_B^-$ exhibits the same feature at the same field. We use same spin-Hamiltonian parameters are taken from Ref. [5].

## §S7. Spin Hamiltonian model

The spin Hamiltonian of the central-spin model is

$$H = H_{ele} + H_{bath} + H_{ele\text{-}bath}, \quad \text{(S4)}$$

where

$$H_{ele} = -\gamma_e \vec{B}_0 \cdot \vec{S} + \vec{S} \cdot \overleftrightarrow{D} \cdot \vec{S}, \quad \text{(S5)}$$

$$H_{bath} = -\vec{B}_0 \cdot \sum_i \gamma_{n_i} \vec{I}_i + H_{dd} + \sum_i \vec{I}_i \cdot \overleftrightarrow{Q}_i \cdot \vec{I}_i, \quad \text{(S6)}$$

$$H_{ele\text{-}bath} = \vec{S} \cdot \sum_i \overleftrightarrow{A}_i \cdot \vec{I}_i. \quad \text{(S7)}$$

Here $\gamma_e$ and $\gamma_n$ are the gyromagnetic ratios of the electron spin and the i-th nuclear spin, $B_0$ is the external magnetic field applied along the c-axis (z direction), D is the electron zero-field splitting (ZFS) tensor (nonzero only for the spin-triplet defects), $Q_i$ is the nuclear quadrupole tensor (active for $I \geq 1$ nuclei), and $A_i$ is the hyperfine tensor of the i-th bath nuclear spin. The nuclear-nuclear dipolar coupling is

$$H_{dd} = \frac{\mu_0}{4\pi} \sum_{i<j} \gamma_{n_i} \gamma_{n_j} \left[ \frac{\vec{I}_i \cdot \vec{I}_j}{r_{ij}^3} - \frac{3\,(\vec{I}_i \cdot \vec{r}_{ij})(\vec{I}_j \cdot \vec{r}_{ij})}{r_{ij}^5} \right], \quad \text{(S8)}$$

and the quadrupole interaction is

$$H_Q = \frac{eQ}{6I(2I-1)} \sum_{\alpha\beta} V_{\alpha\beta} \left[ 3/2\,(I_\alpha I_\beta + I_\beta I_\alpha) - \delta_{\alpha\beta} I^2 \right], \quad \text{(S9)}$$

with $V_{\alpha\beta}$ the electric-field gradient (EFG) tensor, e the elementary charge, and Q the nuclear electric quadrupole moment of the isotope. Throughout we quote the quadrupole coupling constant as $C_Q = eQV_{zz}/h$, using Q = 84.59 mb for $^{10}B$, 40.59 mb for $^{11}B$, and 20.44 mb for $^{14}N$. The isotope $^{15}N$ has I = 1/2 and therefore no quadrupole moment. We adopt the secular approximation in $H_{ele\text{-}bath}$, dropping the $S_x$ and $S_y$ components of the electron-spin operator and retaining the full $S_z \cdot (A_{zx} I_x + A_{zy} I_y + A_{zz} I_z)$ coupling for each bath nucleus. This is justified at the magnetic fields considered here, where the electron Zeeman energy (and, for the spin-triplet defects, also the ZFS) far exceeds the hyperfine couplings.

In addition to the direct dipolar coupling of Eq. (S8), nuclear spins coupled to a common electron spin acquire a second-order hyperfine-mediated interaction through virtual electron-spin flips. To second order in $|A_i| / |\Delta E_{\alpha\beta}|$, the contribution for the qubit in eigenstate $|\alpha\rangle$ takes the form [6, 7]

$$H_{ij}^{(2,\alpha)} = \sum_{\beta\neq\alpha} \frac{(\vec{s}_{\alpha\beta}\cdot\overleftrightarrow{A}_i\cdot\vec{I}_i)(\vec{s}_{\beta\alpha}\cdot\overleftrightarrow{A}_j\cdot\vec{I}_j)+(i\leftrightarrow j)}{E_\alpha - E_\beta}, \quad \text{(S10)}$$

where $s_{\alpha\beta} = \langle\alpha|S|\beta\rangle$ is the matrix element of the electron-spin operator between the qubit eigenstates and $\Delta E_{\alpha\beta} = E_\alpha - E_\beta$ is the relevant electron-spin energy gap (the electron Zeeman energy $\gamma_e B_0$ for spin-doublet defects, or a combination of $\gamma_e B_0$ and the ZFS D for spin-triplet defects).

## §S8. Convergence tests of CCE

All tests use the $C_N$ defect in the $^{11}B^{15}N$ nuclear spin bath as a representative system, evaluated at two magnetic-field values that probe the two distinct decoherence regimes identified in the main text, $B_0$ = 500 G (single-spin-dynamics regime, below the transition boundary) and $B_0$ = 20000 G (pair-spin-dynamics regime, above the transition boundary).

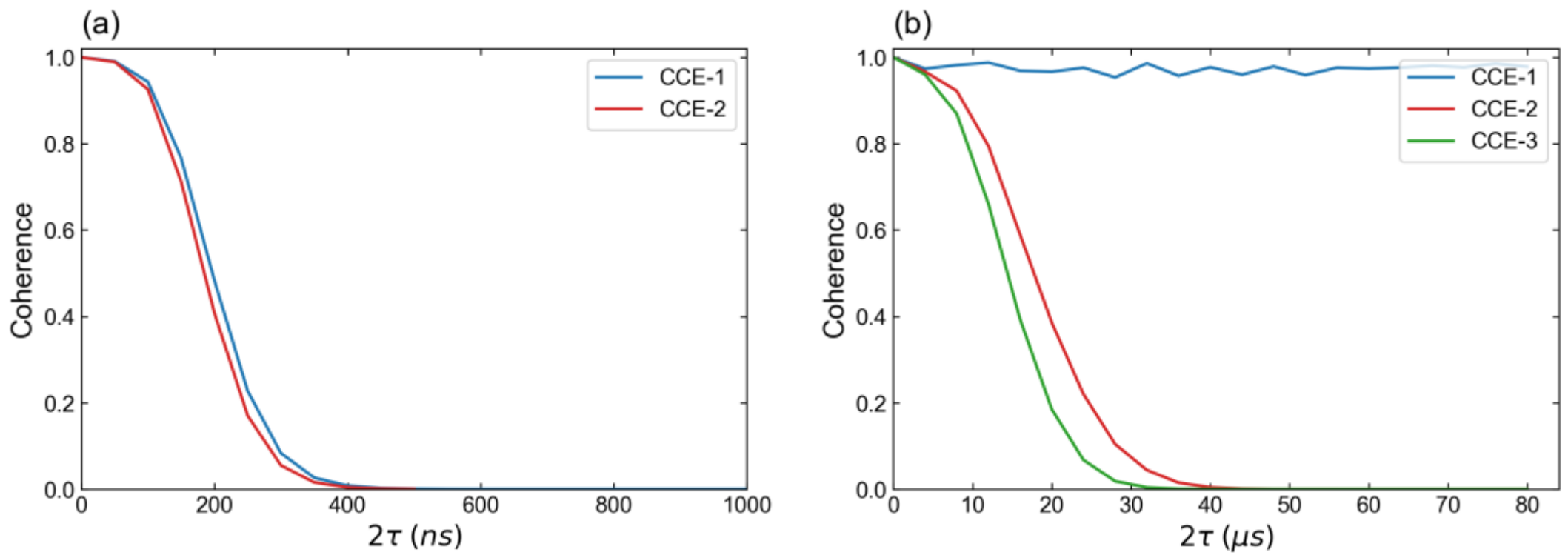

**Figure S10.** Cluster order convergence of the Hahn-echo coherence function for the $C_N$ defect in the $^{11}B^{15}N$ bath. CCE-1 and CCE-2 are compared at (a) $B_0$ = 500 G, and all three orders at (b) $B_0$ = 20000 G.

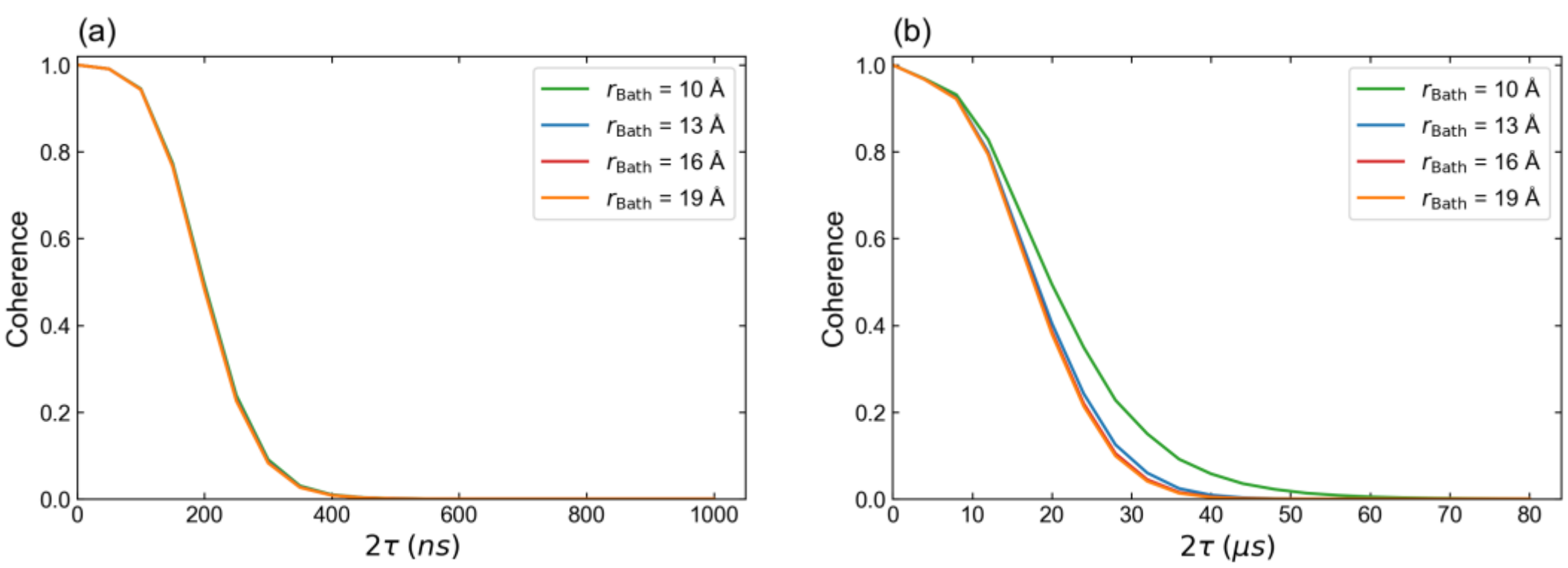

**Figure S11.** Bath-radius ($r_{Bath}$) convergence of the Hahn-echo coherence function for the $C_N$ defect in the $^{11}B^{15}N$ bath. Coherence curves for $r_{Bath}$ = 10, 13, 16, 19 Å are overlaid at (a) $B_0$ = 500 G and (b) $B_0$ = 20000 G, with the dipolar-coupling radius fixed at $r_{Dip}$ = 6 Å.

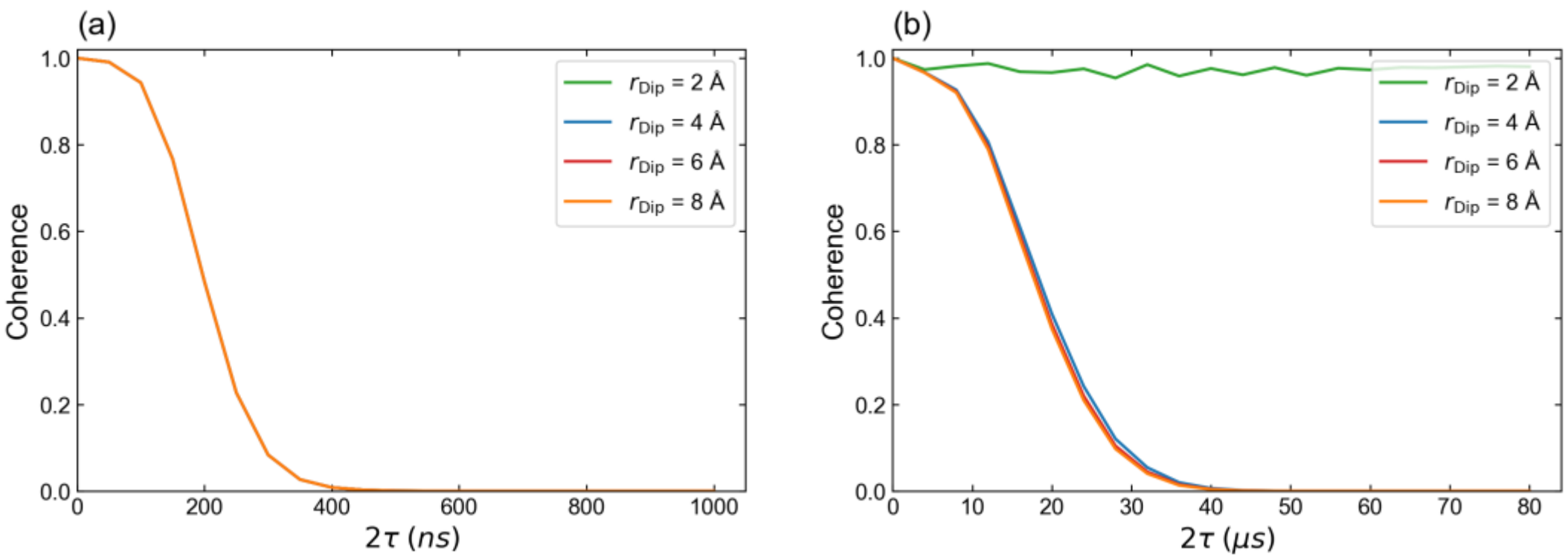

**Figure S12**. Dipolar-coupling-radius ($r_{Dip}$) convergence of the Hahn-echo coherence function for the $C_N$ defect in the $^{11}B^{15}N$ bath. Coherence curves for $r_{Dip}$ = 2, 4, 6, 8 Å are overlaid at (a) $B_0$ = 500 G and (b) $B_0$ = 20000 G, with $r_{Bath}$ fixed at 16 Å. At $B_0$ = 500 G the curves coincide because pair clusters contribute negligibly in the single-spin-dynamics regime, and at $B_0$ = 20000 G the coherence converges by $r_{Dip}$ = 6 Å.

## §S9. Qubit-basis dependence for spin-triplet defects

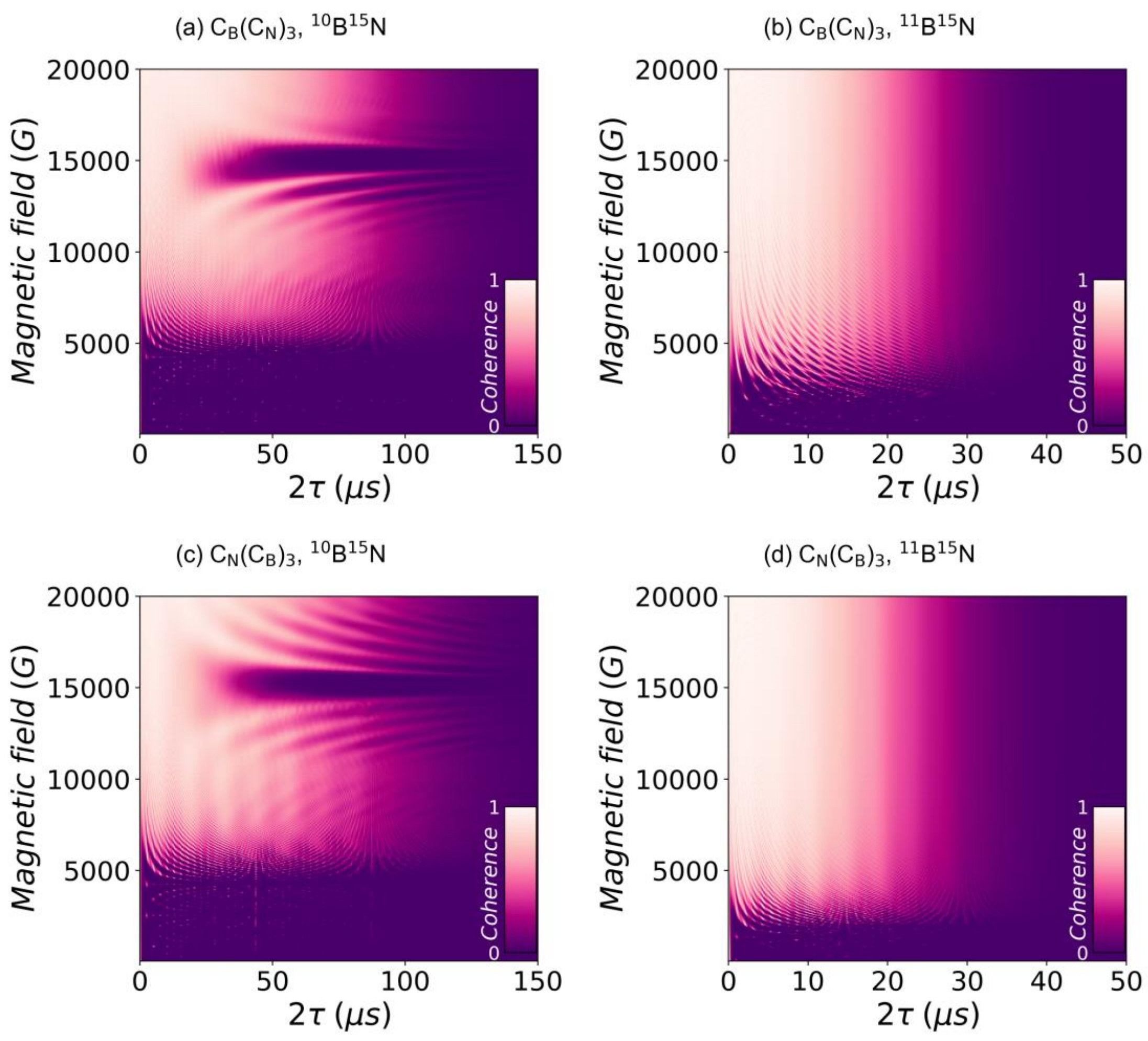

**Figure S13.** Hahn-echo coherence maps of the two spin-triplet defects computed in the $m_S$ = {+1, 0} qubit manifold. (a, b) $C_B(C_N)_3$ and (c, d) $C_N(C_B)_3$, in the $^{10}B^{15}N$ bath (a, c) and the $^{11}B^{15}N$ bath (b, d).